\documentclass{aastex62}
\usepackage{graphicx}
\usepackage{float}

\graphicspath{{./}{figures/}}

\shorttitle{TESS TTV Analysis}
\shortauthors{Pearson}
\begin{document}
\title{A search for multi-planet systems with TESS using a Bayesian N-body retrieval and machine learning}

\correspondingauthor{Kyle A. Pearson}
\email{pearsonk@lpl.arizona.edu}

\author[0000-0002-0000-0002]{Kyle A. Pearson}
\affil{Lunar and Planetary Laboratory \\
University of Arizona \\
1629 East University Boulevard \\
Tucson, AZ, 85721, USA}



\begin{abstract}
Transiting exoplanets in multi-planet systems exhibit non-Keplerian orbits as a result of the gravitational influence from companions which can cause the times and durations of transits to vary. The amplitude and periodicity of the transit time variations (TTV) are characteristic of the perturbing planet's mass and orbit. The objects of interest (TOI) from the Transiting Exoplanet Survey Satellite (TESS) are analyzed in a uniform way to search for TTVs with sectors 1-3 of data. Due to the volume of targets in the TESS candidate list, artificial intelligence is used to expedite the search for planets by vetting non-transit signals prior to characterizing the light curve time series. The residuals of fitting a linear orbit ephemeris are used to search for transit timing variations. The significance of a perturbing planet is assessed by comparing the Bayesian evidence between a linear and non-linear ephemeris, which is based on an N-body simulation. Nested sampling is used to derive posterior distributions for the N-body ephemeris and in order to expedite convergence custom priors are designed using machine learning. A dual input, multi-output convolutional neural network is designed to predict the parameters of a perturbing body given the known parameters and measured perturbation (O-C). \textcolor{black}{There is evidence for 3 new multi-planet candidates (WASP-18, WASP-126, TOI-193) with non-transiting companions using the 2 minute cadence observations from TESS. This approach can be used to identify multi-planet systems and stars in need of longer radial velocity and photometric follow-up than those already performed.}

\end{abstract}


\section{Introduction} \label{sec:intro}

The Transiting Exoplanet Survey Satellite (TESS) is conducting an all-sky photometric survey to discover hundreds of transiting planets around bright stars that are most suitable for mass measurements through radial velocity (RV) observations (\citealt{Ricker2015}; \citealt{Ricker2016}). TESS acquires observations on a 30-minute cadence of all objects in the field of view but increases the cadence to 2-minutes for select bright stars with the goal of detecting small transiting planets \citep{Stassun2018}. Each sector of TESS data is observed for 27.4 days (two spacecraft orbits), allowing the survey to cover most of the sky in 2 years. Simulations have shown that TESS is capable of detecting hundreds of small planets (\citealt{Sullivan2015}; \citealt{Bouma2017}; \citealt{Barclay2018};). To date, TESS has discovered only a few Earth-sized planets (\citealt{Huang2018}; \citealt{Gandolfi2018}; \citealt{Dragomir2018}; \citealt{Vanderspek2019}) and a some larger ones too (\citealt{Nielsen2018}; \citealt{Trifonov2019}; \citealt{Wang2019}; \citealt{Huber2019}; \citealt{Rodriguez2019}). Other discoveries have found new multiplanet systems (\citealt{Leleu2019}; \citealt{Quinn2019}) with only two discoveries yielding measurable transit timing variations so far (\citealt{Kipping2019}; \citealt{Bouma2019}). \citealt{Hadden2018} suggests within the first 2 years of the TESS mission there will be of order 1-10 planet masses precisely measurable with transit timing variations. This dreary estimate contradicts the $\sim$5$\%$ estimate in \cite{Ballard2019} and could be a result of the large constraint on signal-to-noise (7.3) or imperfect knowledge of the underlying distribution of multi-planet system architectures in \citealt{Hadden2018}. 

Transiting exoplanets in multi-planet systems exhibit non-Keplerian orbits as a result of the gravitational influence from companions which can cause the times and durations of transits to vary. Transit timing variations (TTVs) are characteristic of the perturbing planet's mass, eccentricity and period and can be used to constrain the planetary parameters of the perturbing body (\citealt{Agol2005}; \citealt{Holman2005}; \citealt{Nesvorny2008}; \citealt{Agol2018}). Understanding planet-planet perturbations aided in the historical discovery of Neptune by examining the acceleration, then deceleration of Uranus as it passed Neptune (\citealt{Adams1847}; \citealt{Leverrier1877}). Perhaps it is a coincidence but Uranus also exhibits the largest TTV signal of all the Solar System planets (see Table \ref{tab:solarsystem}). 

Past detections of transit timing variations have come mainly from the Kepler mission due to its long term photometric monitoring of the same field of stars for $\sim$4 years \citep{Borucki2010}. Timing variations enable mass and eccentricity measurements of small planets that would otherwise be difficult to measure with radial velocity observations (e.g., \citealt{Jontof-Hutter2016}; \citealt{Hadden2016}). In some cases, the timing variations can be used to assess the presence of additional bodies that may not transit yet still allow for constraints on their mass. Since one of the primary objectives for TESS is to measure masses for 50 planets smaller than 4 R$_{earth}$ timing variations seem like an intriguing pursuit. However, one large constraint on the search for timing variations is TESS's significantly shorter baseline compared to Kepler. Only select areas of the sky\footnote{https://heasarc.gsfc.nasa.gov/docs/tess/primary-science.html} will have data for more than 27.4 days and based on prior studies $\sim$20 transits are generally needed to ensure a unique solution \citep{Nesvorny2008}. This limits the search to planets in compact orbital resonances if they have only 1 sector of data. 

A uniform analysis is conducted on the TESS objects of interest catalog in search of transit timing variations. The two minute cadence data from TESS will help determine prime candidates for RV follow-up that would benefit from mass and orbit constraints. In this paper the observations are described in section 2, followed by the data analysis and transit vetting algorithm in section 3. Section 4 explains the analysis of the light curves and derivation of the linear and non-linear ephemeris models. The results are presented in section 5 and the paper concludes in section 6. 

%Summary of paper sections

\begin{table}
  \caption{TTV Estimates for the Solar System}
  \label{tab:solarsystem}
  \begin{center}
    \leavevmode
    \begin{tabular}{lllll} \hline \hline
    Planet & M$_{planet}$& Period & Eccentricity & TTV$_{max}$  \\ 
           & (Earth) & (day) & & (min) \\
    \hline
Mercury & 0.055 & 87.97 & 0.205 & 1.3  \\ 
Venus & 0.815 & 224.70 & 0.007 & 4.5   \\ 
Earth & 1 & 365.24 & 0.017 & 6.5       \\ 
Mars & 0.107 & 686.98 & 0.094 & 42   \\ 
Jupiter & 317.83 & 4332.59 & 0.049 & 1467 \\  
Saturn & 95.16 & 10759.22 & 0.056 & 5652 \\ 
Uranus & 14.54 & 30588.74 & 0.046 & 11055 \\ 
Neptune & 17.15 & 59799.90 & 0.0113 & 6538 \\
% to do look at derivative of Neptune O-C
    \end{tabular}
  \end{center}
\end{table}

\section{Observations}

\begin{figure}[H]
\includegraphics[scale=0.65]{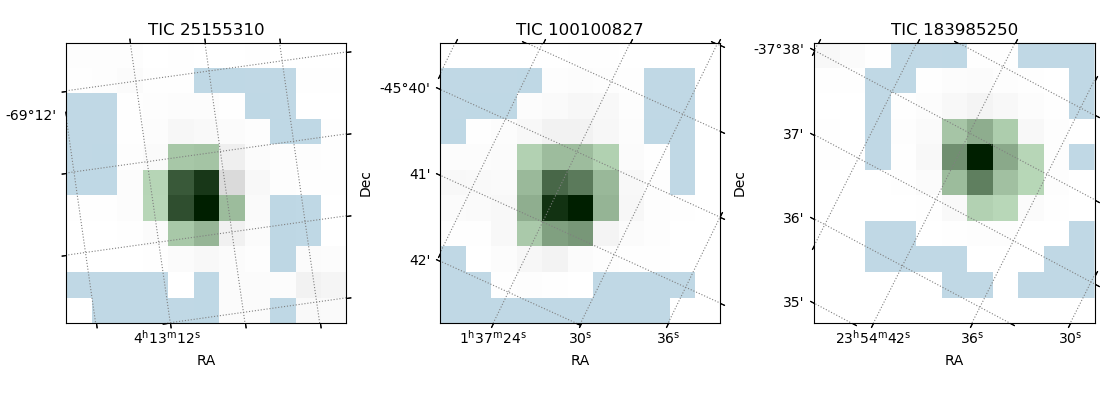}
\hspace{-0.25in}
\caption{ Three stacked \textit{TESS} images of the 11$\times$11 pixel stamp centered around the three targets of interest in this study. The aperture used for background subtraction and for the photometric flux extraction are presented in blue and green, respectively. For reference TIC 25155310 is WASP-126, TIC 100100827 is WASP-18 and TIC 183985250 is also known as TOI 193. %We cross reference the GAIA archive and over plot bright sources near each of our targets, with the target GAIA magnitude in green and the near by sources in red. 
}
\label{TESS_target_pixel}
\end{figure}

The results presented in this paper use data from Sectors 1, 2 and 3 of the \textit{TESS} spacecraft. Light curves of 74 targets from the TOI catalog \citep{Stassun2018} have their time series photometric measurements analyzed. The targets are observed with a 2-minute cadence using a 11$\times$11 pixel subarray centered on the target. The photometric data were processed using the Science Processing Operations Center (SPOC) pipeline \citep{Jenkins2016}, which is based on the predecessor \textit{Kepler} mission pipeline (\citealt{Jenkins2010}; \citealt{Jenkins2017}). Each target is analyzed in the same manner unless otherwise stated. The flux is extracted from the target pixel using apertures centered on the star extending 2--3 pixels in radius. A background subtraction is performed by computing the average background flux from the lower 50$^{th}$ percentile of pixels within an annulus around the target (see Figure \ref{TESS_target_pixel}). The target pixel files include quality flags that indicate when photometric measurements may have been compromised by non-optimal operating conditions on the spacecraft. All flagged data points are removed from the light curves. Additionally, 3 $\sigma$ outliers from a rolling median filter with a bin size of 32 minutes are removed. 

\subsection{Aperture Photometry}

\begin{figure}
\includegraphics[scale=0.7]{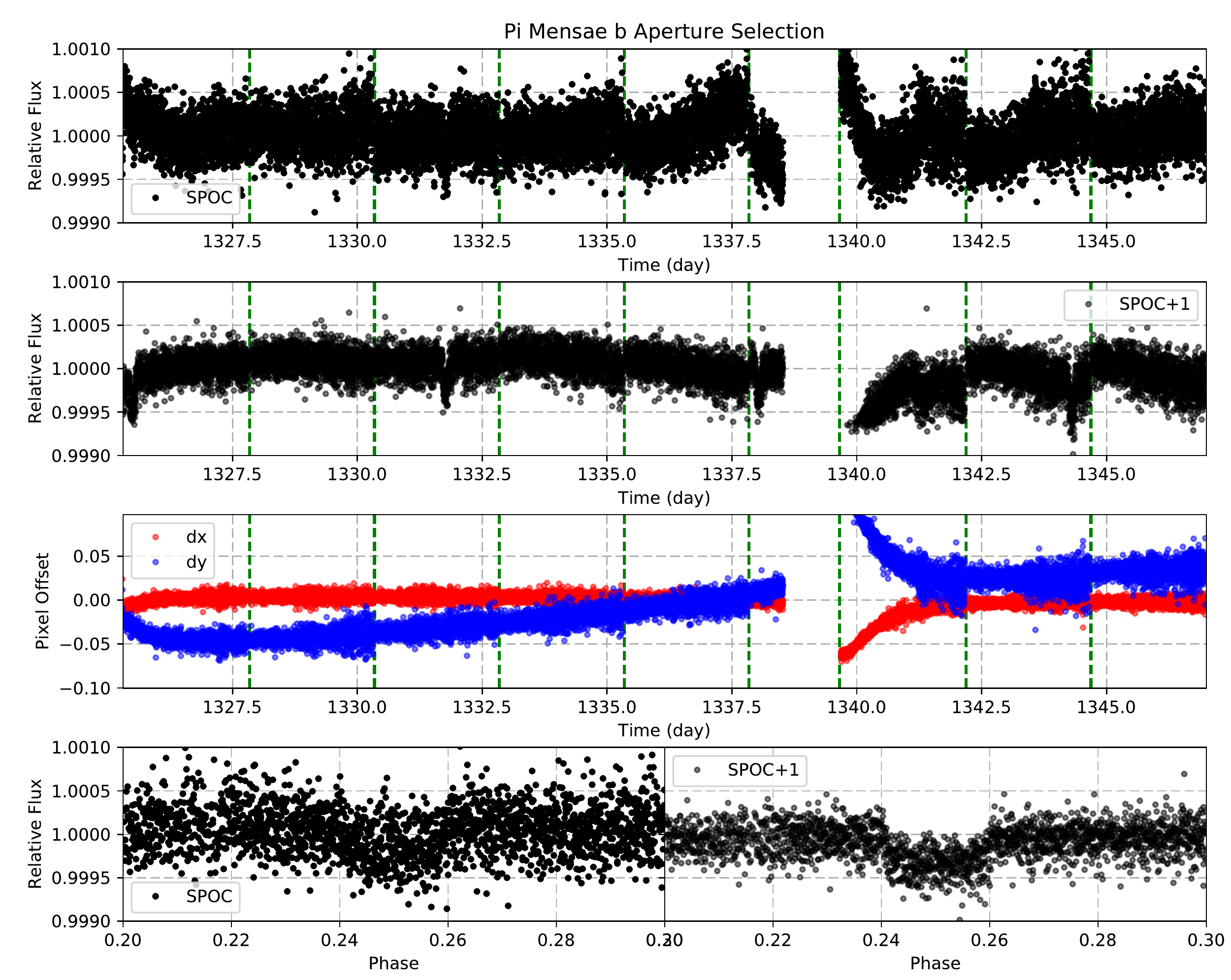}
\caption{ An example of the aperture selection algorithm is validated for the target pi Men b since it was previously studied (\citealt{Huang2018}; \citealt{Gandolfi2018}). The first aperture is produced by the SPOC pipeline (top subplot) and then the SPOC aperture is dilated by 1 pixel indicated as SPOC+1. Flagged measurements are removed prior to performing a 3$\sigma$ clip on the data presented. Additionally, 30 minutes of data before and after the momentum dumps (green dotted line) on the space craft are removed. The gap in the data occurs every 13.8 days on the spacecraft as a result of having to reposition after an orbit. The red and blue dots indicate deviations in the flux weighted centroid of the star (difference from median position). The change in flux with respect to centroid is more correlated to the smaller aperture. The new aperture (SPOC+1) reduces the standard deviation of the residuals by $\sim$30$\%$ without performing any additional pixel level decorrelations. The SPOC+1 aperture yields a transit depth of 285$\pm$4 ppm consistent with the 296$\pm$8 ppm transit depth in the literature \citep{Gandolfi2018}.
}
\label{TESS_pi_men}
\end{figure}

Two different apertures are used to extract the stellar flux and are based on the aperture from the SPOC pipeline. The SPOC aperture is given in the target pixel files and another aperture is a dilated version of the original SPOC aperture. The larger aperture is dilated by one pixel which is used to incorporate more flux from the star and has been found to reduce pointing induced systematics for some stars (see Figure \ref{TESS_pi_men}). The aperture which produces the lowest scatter in the residuals of the phase folded light curve fit is used to model the transit parameters. Figure \ref{TESS_pi_men} shows an example of the aperture selection for the target Pi Mensae b because it has been characterized by two different groups (\citealt{Huang2018}; \citealt{Gandolfi2018};) The original SPOC aperture for Pi Mensae b is a little small and because of that it is sensitive to sub-pixel shifts of the star on the detector. The flux for Pi Men b can vary up to a few hundred ppm with just a 0.1 pixel shift in the centroid. Increasing the aperture size decreased the scatter in the residuals by $\sim$30$\%$ (see Figure \ref{TESS_pi_men}).

\section{Data Analysis}
A custom pipeline is designed to analyze the time series measurements with artificial intelligence being used to vet data that does not exhibit a transit-like shape. \textcolor{black}{A vetting algorithm ensures each target is processed in a homogeneous manner which reduces the removal of arbitrary choices made by human vetting.} Vetting targets saves computational time because full parameter posteriors are derived for each light curve fit using nested sampled ($\sim$10-30 seconds per light curve fit to achieve $\sim$10000 evaluations of the parameter space). \textcolor{black}{Since TESS will observe a plethora of transit events from 1000s of TOIs,} some of which will have low signal, it's crucial to dedicate the time to model each light curve with a global optimization scheme (e.g. nested sampling and not Levenberg-Marquardt) in order to produce robust uncertainties from posteriors. This study focuses only on planets from the TOI catalog since they are prioritized for follow-up observations. Additionally, all existing multi-planet systems are ignored and will be analyzed in the future and contributed to community driven follow-up.

\subsection{Transit Vetting with Artificial Intelligence}
The ideal algorithm for detecting planets should be fast, robust to noise and capable of learning and abstracting highly non-linear systems. A neural network (or deep net) trained to recognize planets with photometric data provides the ideal platform. Deep nets are composed of layers of ``neurons'', each of which are associated with different weights to indicate the importance of one input parameter compared to another. A neural network is designed to make decisions, such as whether or not an observation detects a planet, based on a set of input parameters that treat, e.g. the shape and depths of a light curve, the noise and systematic error, such as star spots. The discriminative nature of this deep net can only make a qualitative assessment of the candidate signal by indicating the likelihood of finding a transit within a subset of the time series. The advantage of a deep net is that it can be trained to identify very subtle features in large data sets. This learning capability is accomplished by algorithms that optimize the weights in such a way as to minimize the difference between the output of the deep net and the expected value from the training data. The network does not rely on hand designed metrics to search for planets, instead it will learn the optimal features necessary to detect a transit signal from training data. Neural networks have been used in a few planetary science applications including those for transit detection and atmospheric characterization (\citealt{Waldmann2016}; \citealt{Kipping2017}; \citealt{Pearson2018}). A notable example used a neural network to validate the detection of a few Kepler planets by assessing the transit-like shape of the data \citep{Shallue2018}. This study adopts a different approach and uses deep nets to vet transit signals prior to fitting the light curves. The extensive TESS target list requires an efficient analysis in order to prioritize follow-up projects. 

A convolutional neural network (CNN) is used to analyze TESS time series data and and it is based on the CNN design in \cite{Pearson2018}. A convolutional neural network (CNN) is well suited for processing time series measurements since the input data is correlated to one another through time. The CNN computes local features in the data by convolving it with multiple filters. Afterwards, the data is down sampled using an average pooling layer which mimics the effect of binning in time to reduce photometric scatter and decreases the number of weights in the next layer. After passing through two convolutional layers the data is processed using 3 layers of neurons which are connected fully. The final CNN network has $\sim$23,000 free parameters \textcolor{black}{(i.e. the weights within all the neurons)} that are optimized from training data. Increasing the number of weights can cause a neural network to over-fit where it learns to classify examples based on features not associated with the signal (e.g. the noise). Therefore, a set of new set of light curves are created only to validate and test the neural network. Optimizing the weights of the CNN is done by training the network on real TESS data with injected transits. Afterwards, a recovery test is performed to examine the sensitivity of detecting transits at low SNR. The CNN was trained using 20000 random light curves with varying SNR where the injected noise was based on the residuals to TESS light curves.

\begin{figure}
\includegraphics[scale=1]{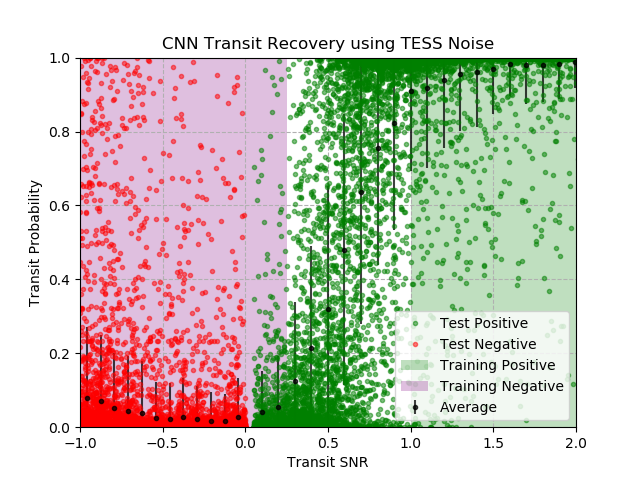}
\caption{ A transit recovery test using a convolutional neural network trained on simulated light curves with TESS residual noise added in. The neural network was trained using 10000 simulated light curves with varying SNR where the injected noise was based on the residuals to the TESS light curve fits. The residuals have the transit signal removed but variations due to instrumental and stellar variability still included. The shaded regions indicate where in the phase space the CNN algorithm was trained. Negative SNR regions indicate the transit shape is inverted. The dots represent data the neural network has not seen before and its predictions. The black points are transit probabilities averaged over bins of 0.2 SNR. The neural network was cautious in selecting transits and has a 15.7$\%$ false negative rate because a high tolerance is used (0.8) for classifying the transit probability. Additionally, the network has a 1.35$\%$ false positive rate only within the conditions tested here and a lower rate for detrended data. This CNN algorithm has an average detection accuracy of 99$\%$ or more for transit depths that are at least 2 $\times$ greater than the noise. }
\label{TESS_recovery}
\end{figure}

The training data is based on TESS residuals which have the transit signal removed but variations due to instrumental and stellar variability still included. \textcolor{black}{Residuals were computed from TOI candidates with SNRs greater than 3 because removing the transit signal is more reliable and less prone to biases from the noise. Light curves with non-detections were avoided because extra processing is required to evaluate if the data includes e.g. a highly variable star, a neighboring and contaminating light source or an instrumental effect. All of which can perturb the data in a manner similar to a transit and thus bias the algorithm during training.} The neural network was trained negatively with data of SNR varying between -1 and 0.25. The negative training refers to the training label being 0, indicating no transit is present in the data. The negative SNR indicates an inverted transit signal meant to simulate a systematic. The network is trained negatively on transits with an SNR between 0 and 0.25 because it decreases the false positive rate due to small systematic effects that decrease flux in a manner similar to a transit (e.g. pointing induced variations, stellar effects). Consequently the positive training samples come from transits with an SNR between 1 and 2, typically these transits are easily visible to the human eye. The neural network is trained on batches of 32 light curves from a sample of 10,000 positive and 10,000 negative samples. Afterwards, the network is tested on data it has not seen before in order to assess the sensitivity of detecting low SNR transits. Figure \ref{TESS_recovery} shows the results of the transit recovery test. The neural network was cautious in selecting transits and has a 15.7$\%$ false negative rate because a high tolerance (0.8) is used for classifying the transit probability. Additionally, the network has a 1.35$\%$ false positive rate which considers inverted transit shapes and other sources of systematics perturbing the light curve. The network exhibits a smaller false positive rate if the data is detrended prior to prediction. The CNN algorithm has an average detection accuracy of 98$\%$ or more for transits that are at least 1.25 $\times$ greater than the noise. For more information on the training and design of the deep net here please see the online material\footnote{\url{https://github.com/pearsonkyle/Exoplanet-Artificial-Intelligence}}. 

% interesting behavior where the neural network registers inverted transit shapes as being positive. 

% Their input data must be detrended aprior (e.g. with a spline or Gaussian process), which if done incorrectly can remove the subtle transit of interest. Additionally, their network uses a max pooling layer which focuses on extracting the brightest data points after the convolutional layer. Therefore, it is necessary to remove hot pixels, cosmic rays and other outliers prior to analysis. In comparison, our network uses an average pooling layer which mimics binning data in time in order to reduce the noise. While a sword is capable of cutting butter quite well, we fancy an algorithm closer to that of a steak knife, effective and precise.
Integrating the convolutional neural network into the analysis is essential to expedite the processing of optimal looking transits that would allow for robust constraints on the system parameters. This vetting algorithm is applied to TESS data by computing a binned phase folded light curve and using the CNN to evaluate the probability a transit is present. If the transit probability is greater than 80$\%$ the light curve will be characterized. The phase folded data is truncated to surround the primary transit with $\pm1$ transit duration before and after ingress, respectively. The data is then interpolated onto a grid of 180 data points because the neural network has a fixed input size of 180. \cite{Pearson2018} found that interpolation of a longer time series did not significantly affect the performance of a CNN for simulated data and Kepler light curves, the same is assumed to be true for TESS. A portion of the TOI candidates that were removed from the analysis exhibit either low SNR transits ($< \sim$1) or had systematic effects in the light curves that would not allow for a precise transit measurement.

% We use the residuals of our phase folded light curve fits to model TESS noise and inject transits of known size into them in order to understand the limitations of our vetting algorithm. Transits were injected into the residuals of our TESS data at SNR values between 0.5 and 2 See Figure X. Our model is subject to finding eclipsing binaries however for our small sample size of jsut the TOI catalog for now we can hand pick targets. Based on our algorithm we find XX TOIs that yield a low transit probability: . A visual inspection of these light curves yields systematic and noise ridden light curves of low SNR objects (usually less than 1) and urge caution for future observations. The input to our neural network is a binned-phase folded light curve. The data are first phase folded and then binned to 180 data points from a time window that is 3* transit duration centered on the mid transit. This allows for ample baseline before and after transit necessary to detect a transit shape. No, preprocessing is required (e.g. splines or gaussian processes) our neural network was trained on systematic ridden data and can detect transits even when. 

% TODO include validation metric, transit inject and signal recovery statistics. 
% SNR vs Transit Detection Probability with and without systematics 

\subsection{Light Curve Model and Nested Sampling}
After the light curves are vetted, the phase folded light curve is characterized and the transit parameters are derived. If the SNR of the phase folded light curve is above 2, the individual transits in the time series are fit in order to derive a new ephemeris. The SNR is defined as the amplitude of the transit depth normalized by the standard deviation of the residuals to the phase folded light curve. In most cases, the SNR is approximately $(rp/rs)^2/\sigma_{res}$ unless the transit is highly inclined in which case the amplitude of the transit depth comes from the transit model itself. The transit parameters are derived by optimizing a model's parameters using observations and nested sampling. The following function below is used to maximize the likelihood of a transit model and non-linear correction simultaneously:

\begin{equation} \label{lcmodel}
F_{obs} = \left( a_{0} + a_{1}*t + a_{2}*t^{2} \right) F_{transit}.
\end{equation}

\noindent
Here $F_{obs}$ is the flux recorded on the detector, $F_{transit}$ is the actual astrophysical signal (i.e. the transit light curve, given by a \cite{Mandel2002} model light curve), $a_{i}$ are quadratic correction coefficients. The quadratic term removes non-linear variations in the light curve that would otherwise bias the mid-transit estimate. As the planet transits in front of the host star, brightness contrasts between the stellar limb to modulate the shape of the transit. A quadratic limb darkening model is used and computed from the tool LDTk and the stellar parameters in Table \ref{tab:stellarpars} \citep{Parviainen2015}. 

\begin{table}[H]
  \caption{Stellar Parameters}
  \label{tab:stellarpars}
  \begin{center}
\begin{tabular}{lllllllll} \hline \hline
TIC ID & T mag &  R$_{s}$ [Sun] & T$_{eff}$ [K] & log(g)& [Fe/H] & u$_{1}$ & u$_{2}$ \\  
\hline
25155310 & 10.56 & 1.27 & 5800 & 4.28 & -0.06 & 0.43 & 0.14 \\
100100827 & 8.83 & 1.23 & 6400 & 4.37 & 0.1 & 0.39 & 0.15 \\
183985250 & 9.10 & 0.99 & 5422 & 4.42 & 0.0 & 0.47 & 0.12 \\
...       &      &      &  ... &      &     &      & ... \\  
\end{tabular}
\\ 
Data are available on GitHub. A portion is shown here for
guidance regarding its form and content
\end{center}
\end{table}

The transit parameters are optimized using the multimodal nested sampling algorithm called MultiNest (\citealt{Skilling2006}; \citealt{Feroz2008}; \citealt{Feroz2009}). MultiNest is a Bayesian inference tool that uses the Monte Carlo strategy of nested sampling to calculate the Bayesian evidence allowing simultaneous parameter estimation and model selection. A nested sampling algorithm is efficient at probing parameter spaces which could potentially contain multiple modes and pronounced degeneracies in high dimensions; a regime in which the convergence for traditional Markov Chain Monte Carlo (MCMC) techniques becomes incredibly slow (\citealt{Skilling2004}; \citealt{Feroz2008}). 
Optimization of the hyperparameters for nested sampling enable an efficient search for the global solution while producing a numerical uncertainty from sampling the posterior distribution. 500 live points are used with an evidence tolerance of 0.1 and a sampling efficiency of 50\% to ensure enough points in the prior space are sampled for convergence and to reveal any multimodal posterior distributions. When fitting the phase folded light curve the following parameters are allowed to vary: $R_{p}/R_{*}$, $a/R_{*}$, $inc$, $T_{mid}$, $a_{0}$, $a_{1}$, $a_{2}$. After the phase folded light curve is characterized, individual transits in the time series are characterized for only $a/R_{*}$, $T_{mid}$, $a_{0}$, $a_{1}$, $a_{2}$ while leaving the other values fixed and adopting them from the phase folded light curve fit (See Table \ref{tab:phasedpars}). 

\begin{figure} 
\hspace{-0.4in} 
\includegraphics[scale=0.4]{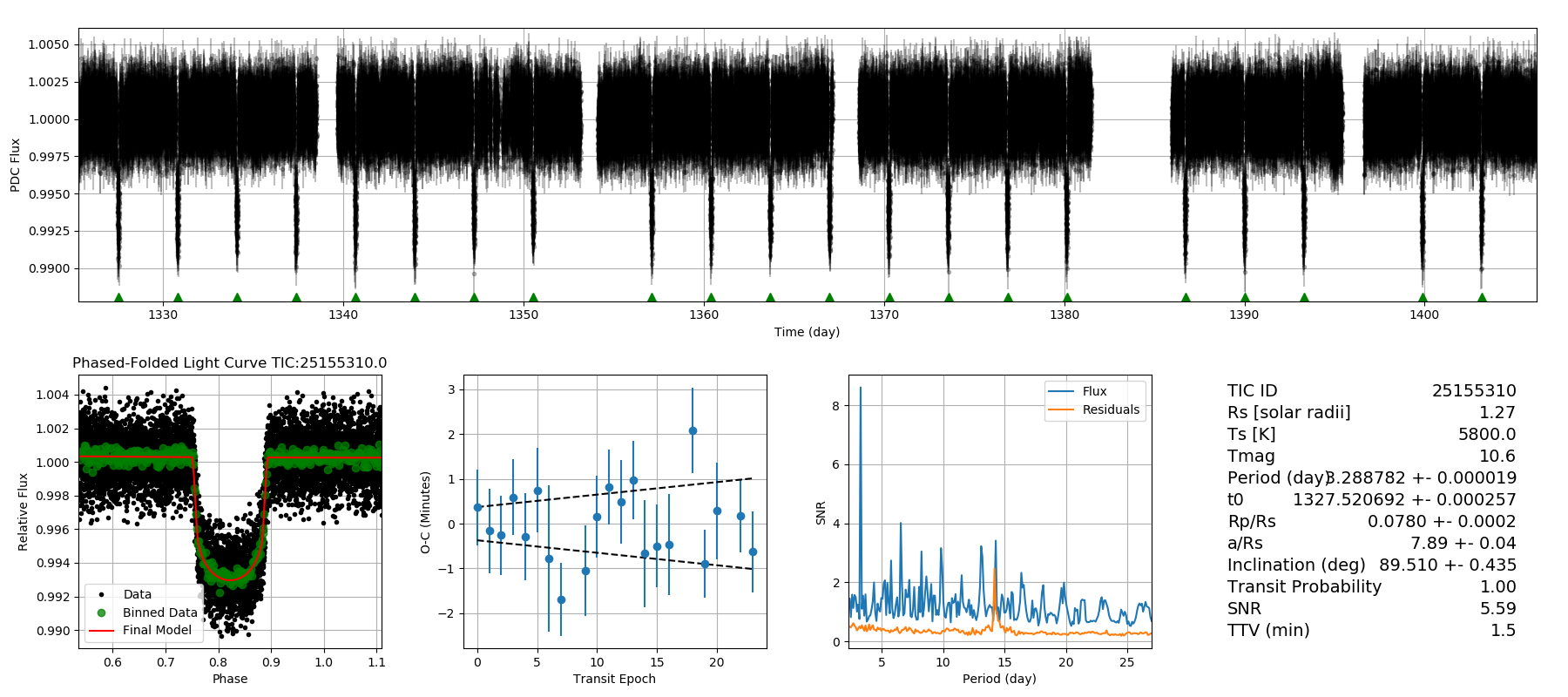}
\caption{ An example data product from the TESS pipeline presented here for the object WASP-126 b. The top subplot shows a full time series for data from sectors 1--3 with the mid transit of each light curve plotted as a green triangle. The bottom left subplot shows a phase folded light curve that has been fit with a transit model to derive the planetary parameters shown in the table on the far right. The green data points are phase folded from the entire time series and binned to a cadence of 2 minutes. The middle subplots show the residuals of a linear ephemeris (calculated) compared to the mid transit times (observed). The dotted line in the O-C plot represents one sigma uncertainties on the linear ephemeris. The middle right subplot shows a transit periodogram for the PDC Flux and for the residuals of the time series, after removing each light curve signal. The spike at 13.8 days in the transit periodogram results from the gaps in the time series due to space craft maneuvers. }
\label{TESS_timeseries_wasp126}
\end{figure}

%\section{Training Data}
%Mimic expected observations from TESS. Create random N-body simulations of 3-body systems with at least 1 transiting exoplanet. We sample stellar masses from 0.5 to 1.5 Msun but later find the stellar mass plays little role in the regression algorithm (see Feature selection section). The inner planet is chosen with a period drawn from a uniform distribution between 1 and 10 days. We only focus on the close-in exoplanets because TESS will acquire the most data from these objects. Additionally, we can limit the degeneracies of the retrieval with at least 20 TTV measurements (consistent with Desvoreny 2008?). The inclination of the first planet is randomly chosen from a distribution between 90 and 90-1.1*inc$_tol$. Where 9$inc_tol$ = tan$^{-1}( (Rs-Rearth)/a)$, a rough approximation for the limiting inclination for a planet to transit. In the event that planet 1 is not transiting, planet 2 (the larger period planet) will be drawn from a conditional dimension such that it has to transit. We use the publicly available Python package called hyperopt to draw our random simulations from because they derive from conditional dimensions. The period of planet 2 is chosen to be at least larger than the hill sphere of planet 1, this prevents unstable orbits which can cause the planet's to undergo orbital evolution far beyond the initial parameters.  
% for training data table say how many transiting planet1,planet2 there are

\subsection{Linear Ephemeris Model}

The orbits of bodies in multi-planet systems are subject to additional gravitational forces than just the star which can cause time-dependent deviations in the orbit parameters. These deviations yield measurable differences in orbital periods and transit durations. For transiting exoplanets, measuring the orbital period is done by fitting a linear function to mid-transit values. The next mid-transit or mid-eclipse time can be calculated using: 
\begin{equation} \label{eq:linear-ephem}
    t_{next} = n_{orbit} * P + T_{mid}
\end{equation}
where $t_{next}$ is the time of the next transit, $P$ is the orbital period of the planet, $T_{mid}$ is the planet's epoch of mid transit and $n_{orbit}$ is the number of orbits that have occurred between $t_{next}$ and $T_{mid}$. Equation \ref{eq:linear-ephem} is used to derive the period and epoch by fitting for those parameters using nested sampling. Discussed more in the previous section, MultiNest is a Bayesian inference tool that uses nested sampling to calculate the Bayesian evidence alongside enabling posterior inference. The Bayesian evidence is compared between a linear and non-linear ephemeris, which accounts for perturbations induced by planetary companions.

\subsection{Non-Linear Ephemeris using N-body simulations}

Planetary companions will perturb the orbits of a transiting planet in a manner indicative to the period and mass of the perturber. To predict the transit times in a system with multiple planets a non-linear ephemeris is computed by: 

\begin{equation} \label{eq:nonlinear-ephem}
    t_{1,next} = f_{Nbody}(n_{orbit},M_{*}, M_{1}, P_{1}, e_{1}, \Omega_{1},  P_{2}, M_{2}, e_{2}, \Omega_{2}) + T_{1,mid}
\end{equation}

where $f_{Nbody}$ is the transit time returned from an N-body simulation, $P_{1}$ is the period of the transiting planet, $P_{2}$ is the period of the companion planet, $M_{2}$ is the mass of the companion planet and $\Omega_{2}$ is the argument of periapsis and $e_{2}$ is the eccentricity of the companion planet and $T_{1,mid}$ is the epoch of mid-transit for the transiting exoplanet.

\begin{figure}
\hspace{-0.2in}
\includegraphics[scale=0.6]{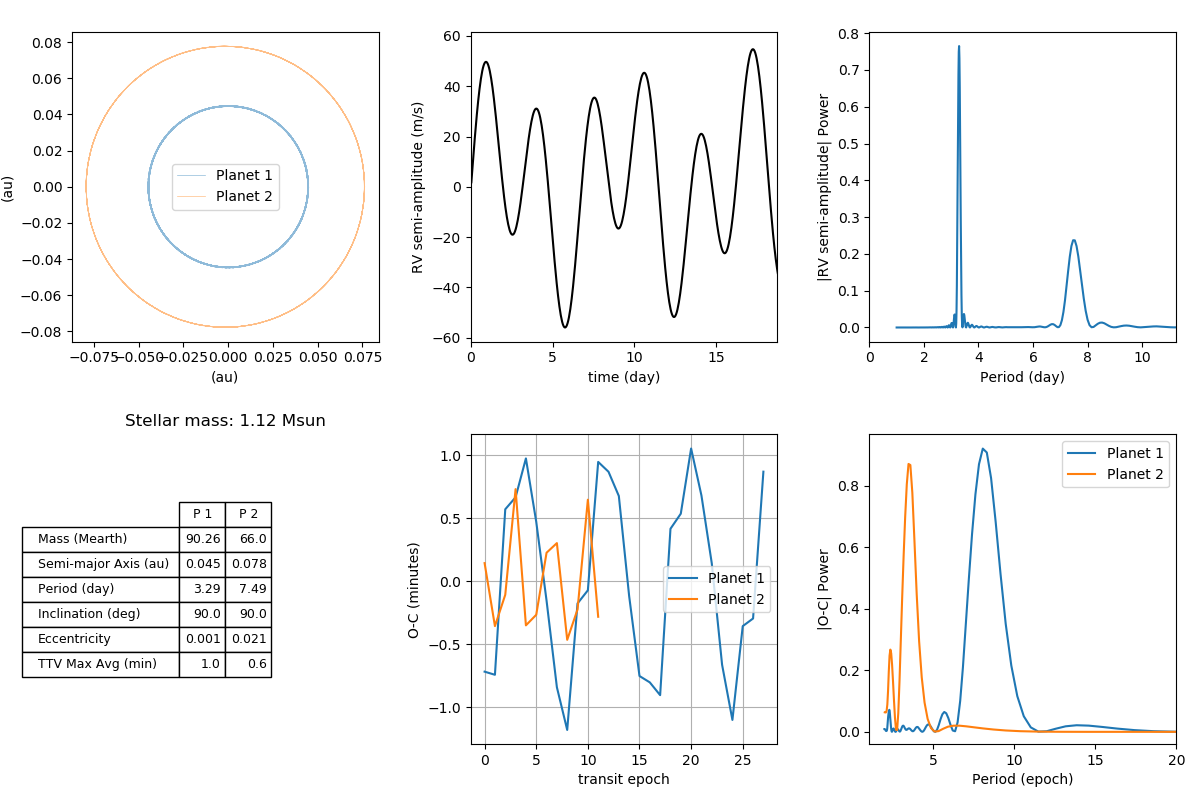}
\caption{ An example data product from one of the N-body simulations. This particular simulation was conducted with a 1.12 M$_{sun}$ star, 90 M$_{Earth}$ planet at 0.045 AU and a 66 M$_{Jup}$ companion at 0.078 AU. \textbf{Top Left}, time series orbit plot of each planet in the system. Each orbit is over plotted and the thicker areas indicate more variation in orbit position. \textbf{Bottom Left}, a table for the system parameters. \textbf{Top Right}, the radial velocity semi-amplitude of the star is plotted as a function of time with a Lomb-scargle periodogram of the RV signal plotted next to it. \textbf{Bottom Right}, a plot showing the deviation from a linear ephemeris fit with linear least squares and a Lomb-scargle periodogram of the O-C data is plotted next to it. }
\label{nbody_sim}
\end{figure}

Computing orbital perturbations on the order of a minute or less requires precise calculations therefore a full N-body simulation is used. REBOUND is an open-source N-body code that features the IAS15 integrator, a 15th-order integrator well suited to simulate gravitational dynamics \citep{Rein2012}. The IAS15 integrator is designed to handle close encounters and high-eccentricity orbits by preserving the symplecticity of Hamiltonian systems using an adaptive time step and a 15th-order modified Runge-Kutta integrator\citep{Rein2015}. 30 minutes is chosen as a default time step for the N-body simulation. A smaller time step of 1 minute was compared against 30 minutes and found to have perturbations within a few seconds of one another. When computing the transit timing variations the position information of the planet is interpolated linearly between time steps to achieve a precision smaller than 30 minutes. Past literature suggests using a time step \textcolor{black}{at least} 1/20 the period therefore 30 minutes is chosen to optimize for speed however it limits the \textcolor{black}{analysis} to transits with orbital periods longer than 10 hours \citep{Nesvorny2008}. An example N-body simulation is shown in Figure \ref{nbody_sim}. 

%  An integrator of this order is chosen because during a parameter retrieval searching portions of the phase space could yield highly dynamic orbits as a result of e.g. interacting hill spheres or unstable orbits.

\begin{figure}
\includegraphics[scale=0.65]{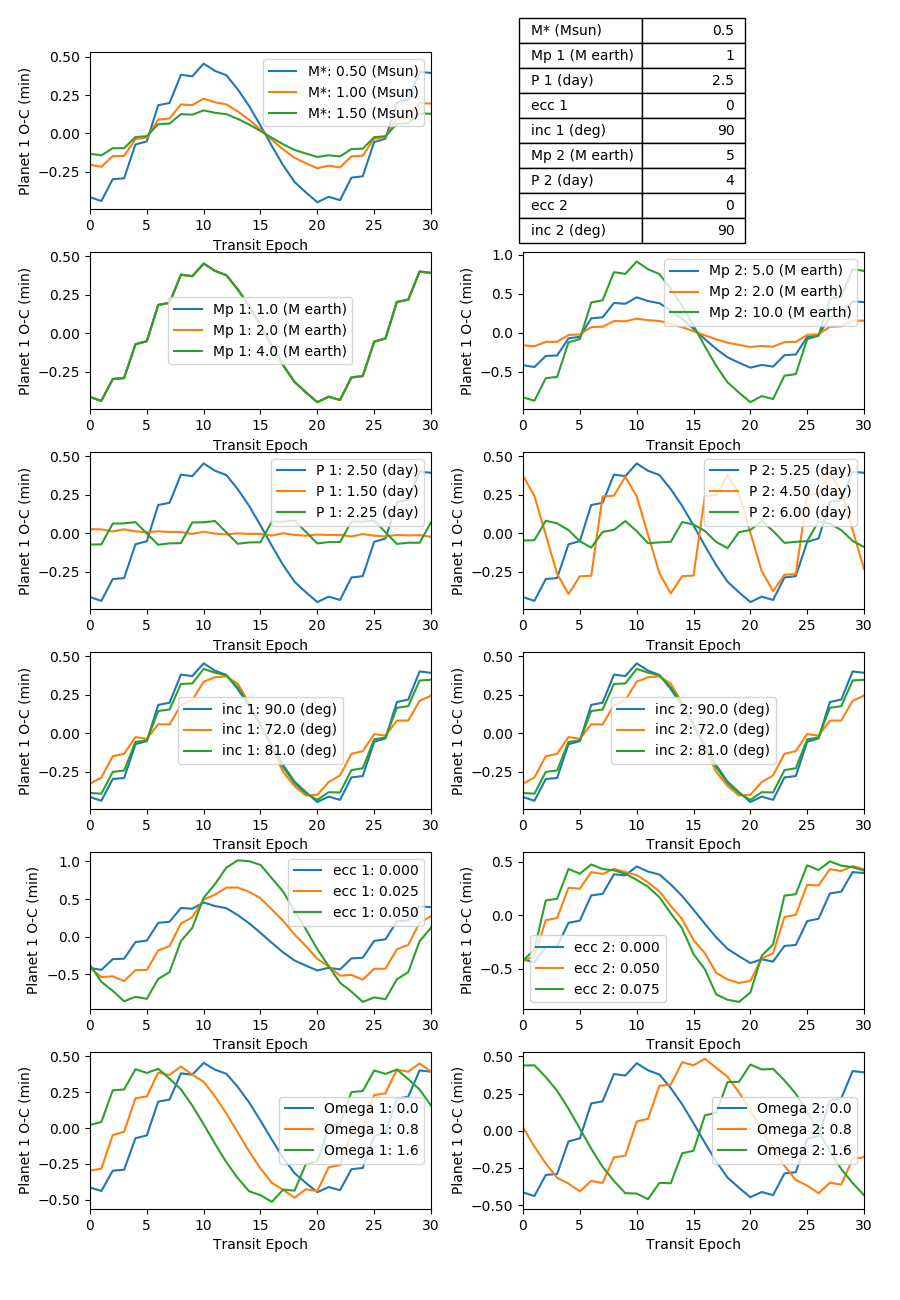}
\caption{ Variations in the TTV signal of just the inner planet as a function of the orbit parameters for the entire system. The default system parameters are in the table at the top and the values in each plot's legend indicate variations from the default parameter set. The inclination of the second planet is difficult to distinguish from the TTV signal. The eccentricity has a degenerate result with mass since they both are proportional to larger deviations. The argument of periastron shifts the TTV signal in epoch and has a smaller effect for orbits with low eccentricity. The values for the argument of periastron are reported in radians. }
\label{nbody_sensitivity}
\end{figure}

N-body simulations contain numerous parameters per planet and the optimization of them can yield millions of combinations to search through. Not every parameter influences the perturbations in a measurable way given the observational precision and therefore certain assumptions can be made about the system to optimize a retrieval. Figure \ref{nbody_sensitivity} shows how each parameter in an N-body simulation perturbs the periodicity of a transiting planet. TTV signals within the phase space tested do not exhibit a strong sensitivity to inclination of either planet. Therefore, the inclination is left as a fixed parameter during an N-body retrieval. The inclination value of the transiting planet is adopted from the phase folded light curve fit while the inclination of the perturbing planet is fixed to 90 degrees. The eccentricity of the perturbing planet is however left as a free parameter. \textcolor{black}{The eccentricity of the inner planet is fixed to the literature value when known or assumed to be 0, yielding an upper limit on the perturbing planet's mass.} The eccentricity of the perturbing planet has an interesting affect on the structure of the TTV signal by making the peaks wider and troughs narrower or vice versa (see Figure \ref{nbody_sensitivity}). There is a degeneracy between the eccentricity of the transiting planet and the mass of a perturbing planet. Essentially, the TTV signal could be from a more massive perturbing planet or an eccentric transiting planet. The easiest way to constrain this degeneracy is by constraining the eccentricity of the transiting planet via secondary eclipse or radial velocity measurement. The N-body retrieval is computed using the Bayesian inference tool Multinest. Typically, the retrieval computes between 5000 and 10000 simulations before converging. Unstable orbits cause the period of the perturber to vary significantly which affects the measured TTV signal up to hours or more. If the solution is unstable, it is returned to the retrieval with a chi-squared value twice as large, this helps constrain the search space for new solutions. 

\begin{figure}
\includegraphics[scale=0.75]{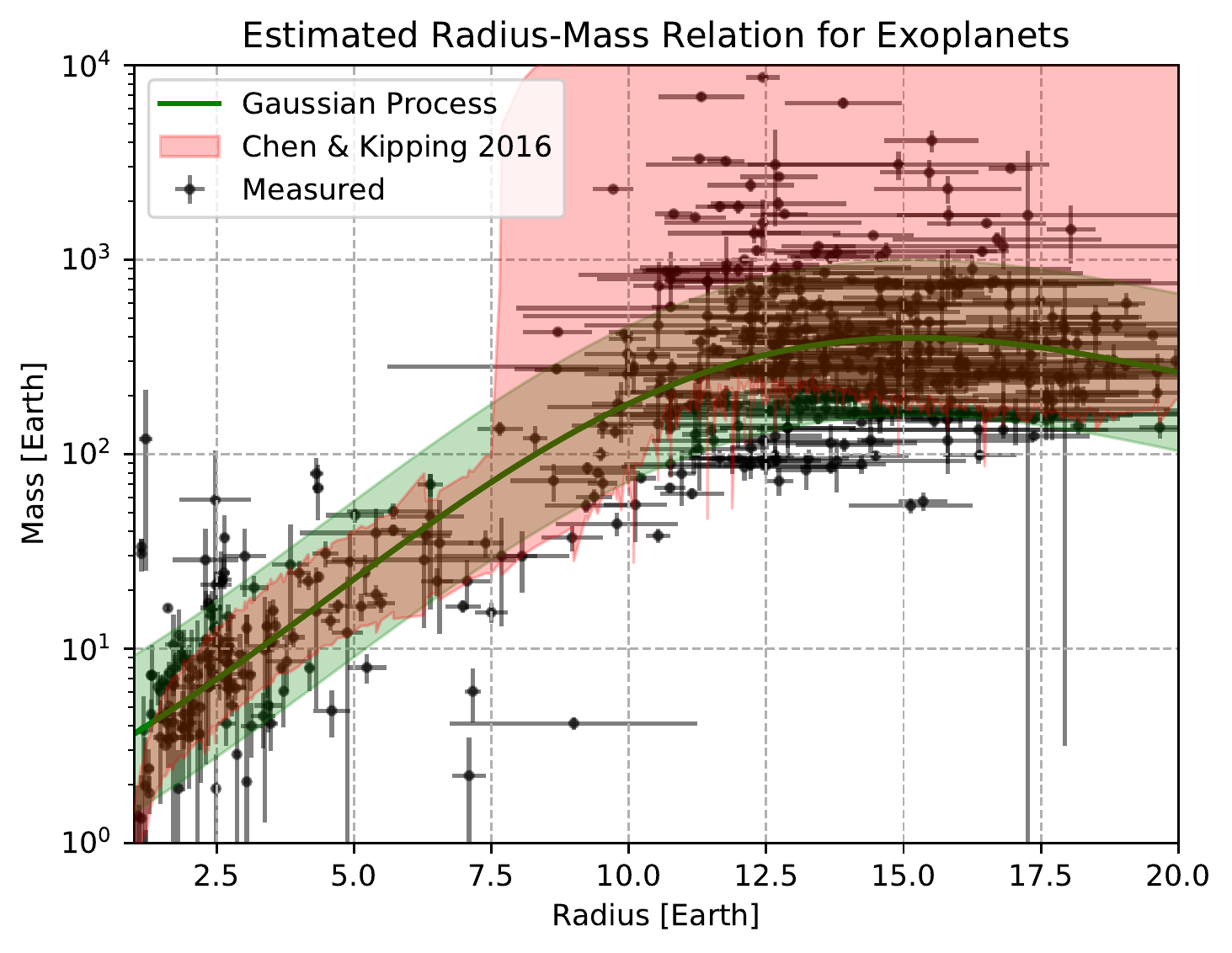}
\caption{ A relationship between planetary radius and mass for all transiting exoplanets with RV and transit measurements in the NASA Exoplanet Archive. A Gaussian Process is used to estimate the correlation between mass and radius as shown in green. The estimator is compared to a probabilistic model using 1000 random samples for each radius value and shown in red \citep{Chen2016}. }
\label{mass_radius}
\end{figure}

\subsection{Estimation of N-body Priors with Machine Learning}

 Matching data with a perturbation model requires testing thousands of N-body simulations in order to find the best-fit parameters and their uncertainties for a perturber. Machine learning can estimate where to search in parameter space before running a more extensive analysis (e.g. using nested sampling). The N-body model depends on at least a few basic parameters; Mass of the star (M$_*$), mass of the inner planet (M$_{1}$), period of the inner planet (P$_1$), mass of the outer planet (M$_2$), period of the outer planet (P$_2$), eccentricity of the outer planet (e$_2$), and argument of periastron for the outer planet (w$_2$). Typically, parameters pertaining to the star and inner planet are known ahead of time. \textcolor{black}{While figure \ref{nbody_sensitivity} shows that TTVs of the transiting planet are not influenced by its mass, that may not hold for non-zero eccentricities or variation of the mass of one order of magnitude (e.g. from super-Earths to Jupiter-like planets), and M$_2$ will be influenced by the choice of M$_1$. When the mass of M$_1$ is not known aprior it is estimated using an empirical mass radius relation. The empirical mass radius relation uses all of the transiting exoplanets with mass measurements from the NASA Exoplanet Archive. A Gaussian Process is used to estimate a general trend between mass and radius (see Figure \ref{mass_radius}). }
 
 A machine learning algorithm is designed to predict the parameters of the perturbing body (M$_2$, P$_2$, e$_2$, w$_2$) given the known parameters (M$_*$, M$_1$, P$_1$, e$_1$, w$_1$) and measured perturbation (i.e. measured mid-transit values subtracted from a linear fit). To do this a dual-input, multi-output regression model is created. The independent features (M$_*$, M$_1$, P$_1$) are analyzed using a fully connected neural network while the time-dependent features are analyzed using a convolutional neural network. The output of each branch is piped into a fully connected neural network and finally, 4 parameters are predicted representing the planetary and orbital elements of the perturbing body. 
 
Training data is simulated in order to optimize the neural network for analyzing high cadence data from TESS observations. The TESS mission is conducting an all sky survey and will observe portions of the sky for as many days at 180 but also as little as 27 depending on the RA and Dec \footnote{https://heasarc.gsfc.nasa.gov/docs/tess/operations.html}. Therefore random simulations of short-period planets around solar type stars are created to train the neural network. Stellar masses range between 0.85 and 1.15 solar since the targets of interest in this study are around solar type stars. Inner planets are generated from a uniform distribution between 0.66-150 M$_{Earth}$ and a period between 0.75 and 10 days. The outer planet is generated from a conditional probability distribution in order to avoid randomly generating unstable orbits. The outer planet is randomly generated with a mass ratio between 0.25 and 3 with respect to the inner planet. The period of the outer planet is generated with a period ratio between 1.25 to 2.5 but will be larger if the random period is within the hill sphere of the first planet. One last constraint is in place to prevent generating unstable orbits; only simulations producing TTVs less than 10 minutes are used. 

\begin{figure}[H]
\centering
\textbf{Distribution of Neural Network Uncertainty}

\includegraphics[scale=0.8]{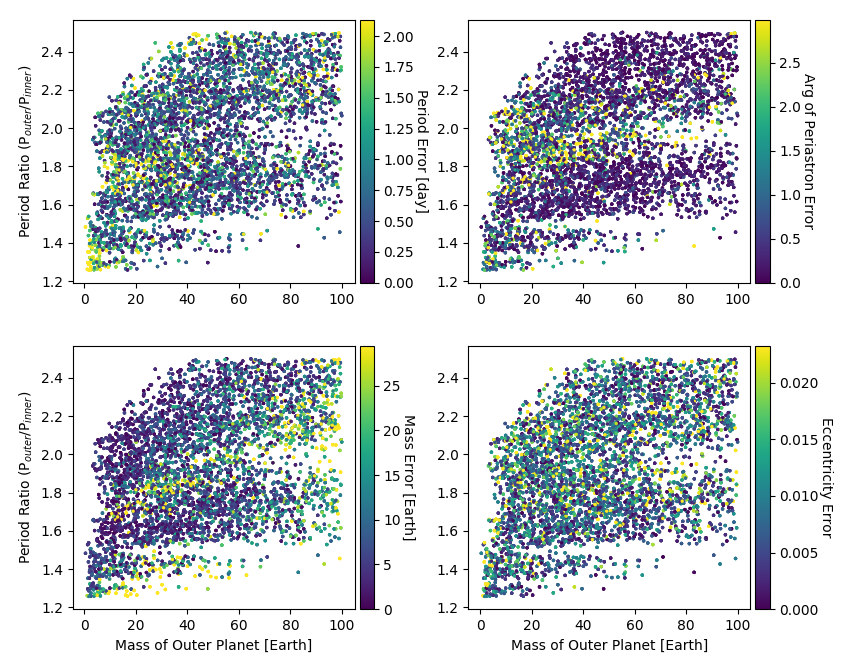}
\caption{ A neural network is trained to make a prediction on the size of the perturbing body given the stellar and inner planet parameters and the measured TTV. Uncertainties in each parameter (M$_{2}$,P$_{2}$, e$_{2}$, $\Omega_{2}$) predicted by a neural network on test data it was not trained on.  An interpolation between the uncertainties is used to estimate priors in an N-body retrieval. The priors are constructed using a uniform distribution centered on the prediction and spanning +/-2$\times$ the error.}
\label{nn_error_dist}
\end{figure}

\begin{figure}[H]
\centering
\textbf{Neural Network Prediction of Orbit Priors}

\includegraphics[scale=0.35]{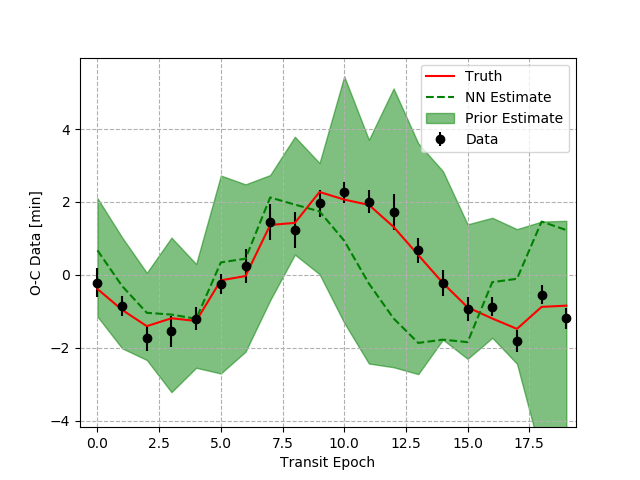}
\includegraphics[scale=0.35]{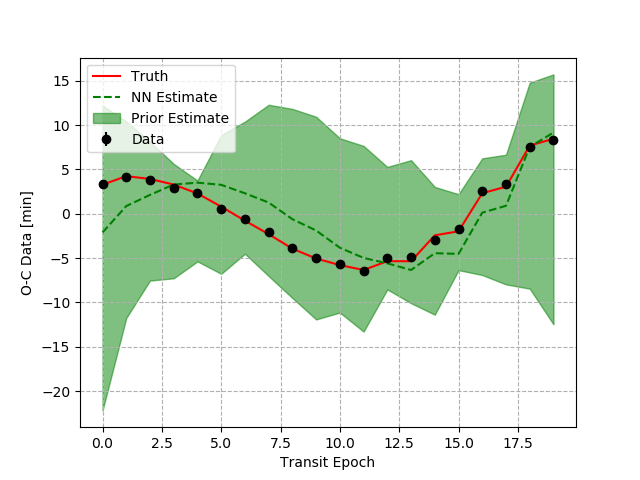}
\includegraphics[scale=0.35]{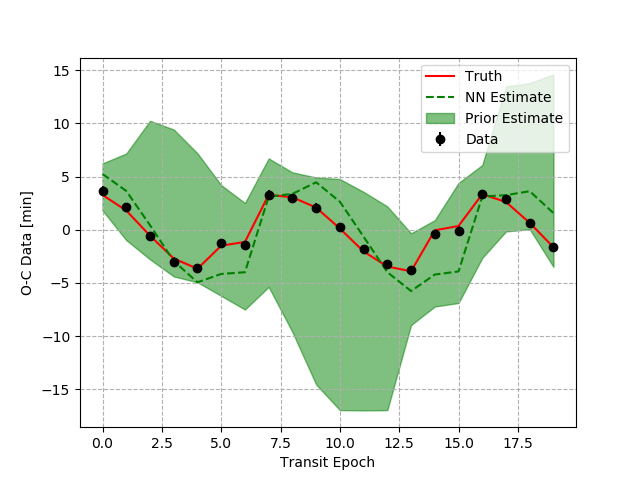}
\caption{ Estimates from the neural network regressor are compared against N-body simulations it wasn't trained with. The min and max bound of a uniform prior is estimated using $\pm$2 $\sigma$ uncertainties from the neural network error in Figure \ref{nn_error_dist}. The priors are then used to fit the transit times using nested sampling and an N-body model. }
\label{nn_prior}
\end{figure}

The constraints on the training data are tailored for this specific study but the code can be easily modified to account for larger TTVs or smaller mass stars. Two variations are made to the training data in order to improve the network's accuracy during training. \textcolor{black}{For every parameter combination (M$_2$, P$_2$, e$_2$) the argument of periastron is uniformly distributed between -$\pi$ and +$\pi$ with 16 points.} Additionally, for each parameter combination the O-C data is varied according to a Gaussian distribution with a standard deviation 10$\%$ of the TTV amplitude. Varying the O-C data simulates acquiring different observations. The neural network is trained with 100,000 samples and 100 epochs using the mean squared error as the loss function. After training the neural network, the error and range of reliability are characterized as a function of orbital period ratio and companion mass (See figure \ref{nn_error_dist}). \textcolor{black}{The neural network errors are used to estimate the priors in a MultiNest Fit (described in section 3.4) by extending the errors $\pm$2 sigma (see Figure \ref{nn_prior}).
}

%The machine learning algorithm uses simulations to learn the parameters of interest, in particular (M$_{2}$,P$_{2}$, e$_{2}$, $\Omega_{2}$). The O-C data is randomly perturbed by a Gaussian function with a standard deviation 10$\%$ of the maximum amplitude of the TTV signal. It's important to perturb the data to prevent the neural network from strictly memorizing the structure of the data and it improves the network's ability to generalize. Each simulation is created with 10 different argument of periastron values for every combination of simulation (Ms, M1, P1, M2, P2, e2). 

%The priors for the companion (M$_{2}$,P$_{2}$, e$_{2}$, $\Omega_{2}$) are estimated with a neural network before the retrieval (see Figure \ref{nn_prior}). Retrieving orbit parameters and uncertainties typically takes $\sim$5000 N-body simulations or more. N-body simulations can take up to a few minutes to run depending on how long the integration is. Machine learning is leveraged to expedite the retrieval by estimating the orbit parameters and constraining the prior parameter distribution. Constraining the prior means the retrieval takes less time to run. Specifically, a fully connected neural network is used to map O-C data along with M$_{s}$, M$_{1}$, P$_{1}$ to the orbit parameters of a companion; M$_{2}$, P$_{2}$, e$_{2}$ and $\Omega_{2}$.

% The unstable solutions consistent of planets interacting hill spheres or immense gravitational potentials that fling planets from orbit. We check that the of the period of the planet's orbit does not vary by more than 50--150$\%$ the max TTV signal found in the data. This constrains the parameter space rather quickly and expedites convergence.

%The mass of the transiting planet does not affect the O-C data for itself therefore we adopt values from the RV literature or assume a value from a Gaussian Process trained on Kepler mass and radius measurements for transiting planets with RV data. 

%TESS is performing an all sky survey and will have the opportunity to follow-up existing multi-planet systems in order to constrain ephemerides and place constraints on planetary parameters like mass and period. The transit timing variation of existing multi-planet systems is estimated using the N-body code and reported in Table \ref{tab:ttvestimates}. This subset of planets are ranked by the maximum TTV amplitude from 20 transit epochs. Only planets with periods less than 60 days are used because of the minimum TESS observing baseline is 27.4 days per sector. The system parameters are taken from the NASA Exoplanet Archive and references there in \citep{Akeson2013}. These targets, particularly the ones with TTV amplitudes less than 30 minutes, could benefit from high cadence observations at 2 minutes in order to constrain the ephemeris and perturbing planet's mass.

\section{Results and Discussion}
\begin{figure}
\textbf{Measured TESS Photometric Precision}

\includegraphics[scale=0.75]{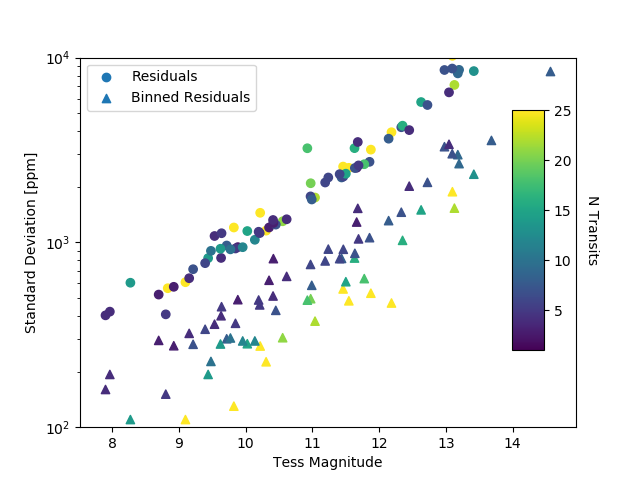}
\caption{ The photometric variability of 74 planet hosting stars from the TOI catalog is quantified in the plot as the standard deviation of the phase folded transit residual. The machine learning algorithm was used to vet non-transit like candidates prior to fitting light curves. The data is from sectors 1-3 and a two minute cadence is used to bin the phase folded data before fitting the binned residual model. TESS reaches an average photometric precision of $\sim$1000 ppm for stars between 9--11 T mag. The TESS photometric precision constrains the ability to detect individual transits of Earth-sized planets at a SNR of 1 to stars with radii less than 0.3 R$_{Sun}$. For stars larger than 0.3 R$_{Sun}$ an Earth sized planet transit would be smaller than 1000 ppm suggesting multiple measurements are needed to detect the planet signal beneath the noise.
}
\label{tess_precision}
\end{figure}

Searching through the TESS objects of interest (TOI) catalog yielded information about the planetary ephemerides and transit parameters of 74 targets. A subset of the catalog is studied for single planet systems with high cadence data, more than 3 orbits, transits longer than 30 minutes, transit probability greater than 80$\%$ from a classifying neural network \textcolor{black}{(see Table \ref{tab:ai})} and a transit SNR greater than at least 2. These constraints enable high precision measurements of multiple transit events which are used to search for planets by measuring transit timing variations. \textcolor{black}{148 different targets entered the pipeline, 30 failed the vetting algorithm and of the 118 left only 74 had an SNR greater than 2 and at least 3 transit measurements.} Figure \ref{tess_precision} shows the photometric variability of TESS stars in this sample as a function of their magnitude. A time series analysis is conducted to derive a new ephemeris for each target. Ephemeris residuals that exhibit periodic structure undergo an in depth analysis to explore perturbing bodies. A neural network is used to estimate the priors for a retrieval and a nested sampling algorithm derives posteriors for an N-body model.

\begin{table}[H]
  \caption{Results of AI Vetting Algorithm}
  \label{tab:ai}
  \begin{center}
\begin{tabular}{ll} \hline \hline
TIC ID & Transit Probability \\  
\hline
  120916706 &                 1 \\
  122612091 &                 1 \\
  159951311 &                 1 \\
  164767175 &         0.0068   \\
  166836920 &                 1 \\
  219253008 &                 1 \\
  220479565 &         0.0016 \\
...     & ... \\  
\end{tabular}
\\ 
Data are available on GitHub. 
\\ A portion is shown here for
guidance regarding its form and content
\end{center}
\end{table}

\begin{table}
  \caption{Phase folded light curve parameters}
  \label{tab:phasedpars}
  \begin{center}
\begin{tabular}{llll} \hline \hline
& WASP-18 b & TOI 193  & WASP-126 b \\
\hline
TIC ID & 100100827 & 183985250 & 25155310 \\ 
Period [day] & 0.9415 $\pm$ 1.34e-06 &  0.7921 $\pm$ 1.80e-05 & 3.2888 $\pm$ 1.94e-05 \\
T$_{mid}$ & 1354.4579 $\pm$ 4.15e-05 & 1354.2151 $\pm$ 3.56e-04 &  1327.5207 $\pm$ 2.57e-04 \\
R$_{p}$/R$_{s}$ & 0.0964 $\pm$ 1.08e-04 & 0.0451 $\pm$ 5.67e-04 & 0.0780 $\pm$ 2.17e-04 \\
a/R$_{s}$  & 3.570 $\pm$ 1.55e-02 & 3.671 $\pm$ 1.27e-01 & 7.887 $\pm$ 3.98e-02 \\
Inclination &  85.68 $\pm$ 2.79e-01 & 75.40 $\pm$ 6.12e-01 & 89.51 $\pm$ 4.35e-01 \\
$\sigma_{res}$ [ppm] & 565 & 611 & 1302 \\
SNR        &  19.16 & 2.65 & 5.59 \\
TTV$_{max}$ (min)   & 0.4 & 3.0 &  1.5 \\
Transits & 47 & 32 & 21 \\
%Sectors &  2 &  1 &  3 \\
\end{tabular}
\end{center}
\end{table}

%The eccentricity of two transiting planets, WASP-18 b and WASP-126 b, are both set to 0 (\citealt{Nymeyer2011}; \citealt{Maxted2016}).

\subsection{WASP-126}
The first target of interest is WASP-126 because it shows subtle periodic structure in the residuals of the linear ephemeris. WASP-126 b is a 0.28 M$_{Jup}$ mass planet at an orbital period of 3.28 day around a G2 5800K star \citep{Maxted2016}. Not much is known beyond the initial discovery of WASP-126 b however it does have a large scale height ($\sim$650 km) and would be an interesting candidate for atmospheric characterization. The best fit parameters for the phase folded light curve are reported in Table \ref{tab:phasedpars}. The time series analysis of WASP-126 b is shown in figure \ref{TESS_timeseries_wasp126}.

\textcolor{black}{
A search for contaminating light sources affecting the transit depth is performed by cross-referencing the TIC catalog for the position and magnitude of neighboring stars. The plate scale of TESS is 21''x21'' which corresponds to 50$\%$ and 90$\%$ of a star’s flux being contained within a 1×1 and 4×4 pixel region around the centroid, respectively \citep{Ricker2015}. The large pixel scale helps TESS in their all sky survey of transiting planets but limits the ability to spatially resolve stars therefore pixels may contain one or more light sources. WASP-126 (T = 10.55) has two stars within an arc minute, TIC 25155316 (T = 14.14) and TIC 25155311 (T = 16.06) at a distance of 26.5'' and 7.9'' respectively. Compared to WASP-126 the nearby sources are 27 and 160 times fainter, respectively. Given the proximity to WASP-126 these objects contribute $\sim$1$\%$ of the star's flux within the aperture used which is comparable to the transit signal of $\sim$0.6$\%$. It is unlikely the transit signal is a systematic effect induced by contamination because the stars would have to completely disappear or move out of the aperture in order to simulate the transit. The stars are at most 1.26 pixels away from the target and the aperture size is roughly 3 pixels in radius. Since the median displacement of the centroid in time is 0.02 pixels with the max being 0.05 pixels, it unlikely the transit signal is being modulated by pointing induced systematics from the contaminating stars. }

\begin{figure}[H]
\centering
\textbf{WASP-126 Posterior for Non-linear Ephemeris}
\includegraphics[scale=0.5]{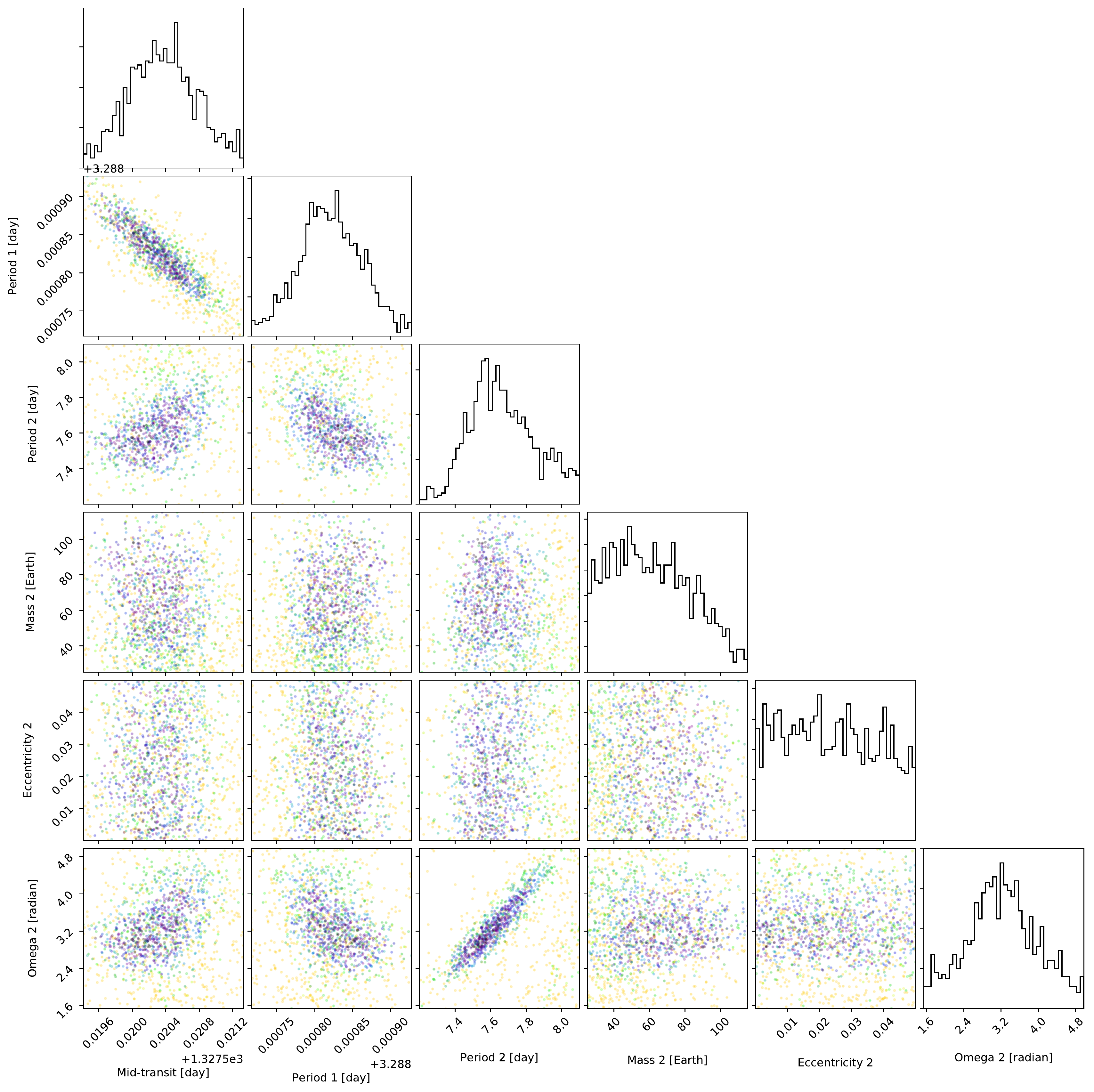}\newline
\caption{ Posteriors for the non-linear ephemeris model are plotted for the WASP-126 system. The color of the data points are coordinated to the fit percentile, with colors darker than yellow indicating fits better than the 50th percentile. The final parameters and uncertainties are in Table \ref{tab:ephem}. The best-fit ephemeris model for WASP-126 is plotted in Figure \ref{wasp126_model}. 
}
\label{wasp126_posterior}
\end{figure}

The transit times for WASP-126 b show periodic structure (see O-C data in figure \ref{TESS_timeseries_wasp126}) and to test the significance of a perturbing planet the data are analyzed with an N-body retrieval described in Section 3.4. The posteriors from the N-body analysis are shown in Figure \ref{wasp126_posterior}. The Bayesian evidence for a linear ephemeris is -18.9 and the Bayesian evidence for a non-linear ephemeris is -15.9, suggesting a non-linear ephemeris is a better fit. The greatest evidence is for a perturbing planet at an orbital period of 7.65 $\pm$ 0.27 days with a mass of 64.0$\pm$26.9 M$_{Earth}$. The data is weakly sensitive to the eccentricity of the perturbing body, which is estimated to be \textcolor{black}{$<$ 0.05} (see Table \ref{tab:ephem}). The upper limit on the prior was determined by the machine learning algorithm discussed in Section 3.5. The perturbing body should be detectable with future RV observations and it corresponds to an RV semi-amplitude of $\sim$11--27 m/s. 

There is no evidence for another transiting companion in the WASP-126 system (see Figure \ref{TESS_timeseries_wasp126}). A transit periodogram analysis is conducted on the time series after removing the primary transit signal from every epoch. The large spike in the periodogram is a result of motions from the space craft and seen in every target analyzed, sadly it's not a planet signal. In order for the planet candidate, WASP-126 c, to not transit it must have an inclination less than $\sim$85.5. The inclination constraint places a lower limit on the mutual inclination of the system to be 4 degrees.  

% \begin{figure}[H]
% \includegraphics[scale=0.75]{wasp126_nbody_posterior.png}
% \caption{The posterior parameter distribution of a non-linear ephemeris model for the WASP-126 system computed with an N-body simulation and nested sampling. The period and 0th mid transit are fit for WASP-126 b while a planetary companion is injected into the system with Mass 2 and Period 2. The histograms for the planetary companion show two different modes, a local and global minimum, and because of the degeneracies each mode must be explored separately. The points are color coordinated according to log likelihood values. Values in black represent models with log likelihood values greater than the 35th percentile of all fits, the green points are between the 10th and 35th percentile and the points in red indicate the best fits in the top 10th percentile. With colors the two modes are easily distinguishable in the parameter space with one being at $\sim$600 M$_{Earth}$, $\sim$11 day and another at $\sim$70 M$_{Earth}$, $\sim$7.5 day. The N-body retrieval is run a second time in order to derive uncertainties and the final values between 1-150 M$_{Earth}$ and 5-8.5 day period (see Figure \ref{nested_post_zoom}). }
% \label{nested_post}
% \end{figure}

\begin{figure}[H]
\centering
\textbf{WASP-126 Ephemeris Comparison}

\includegraphics[scale=0.75]{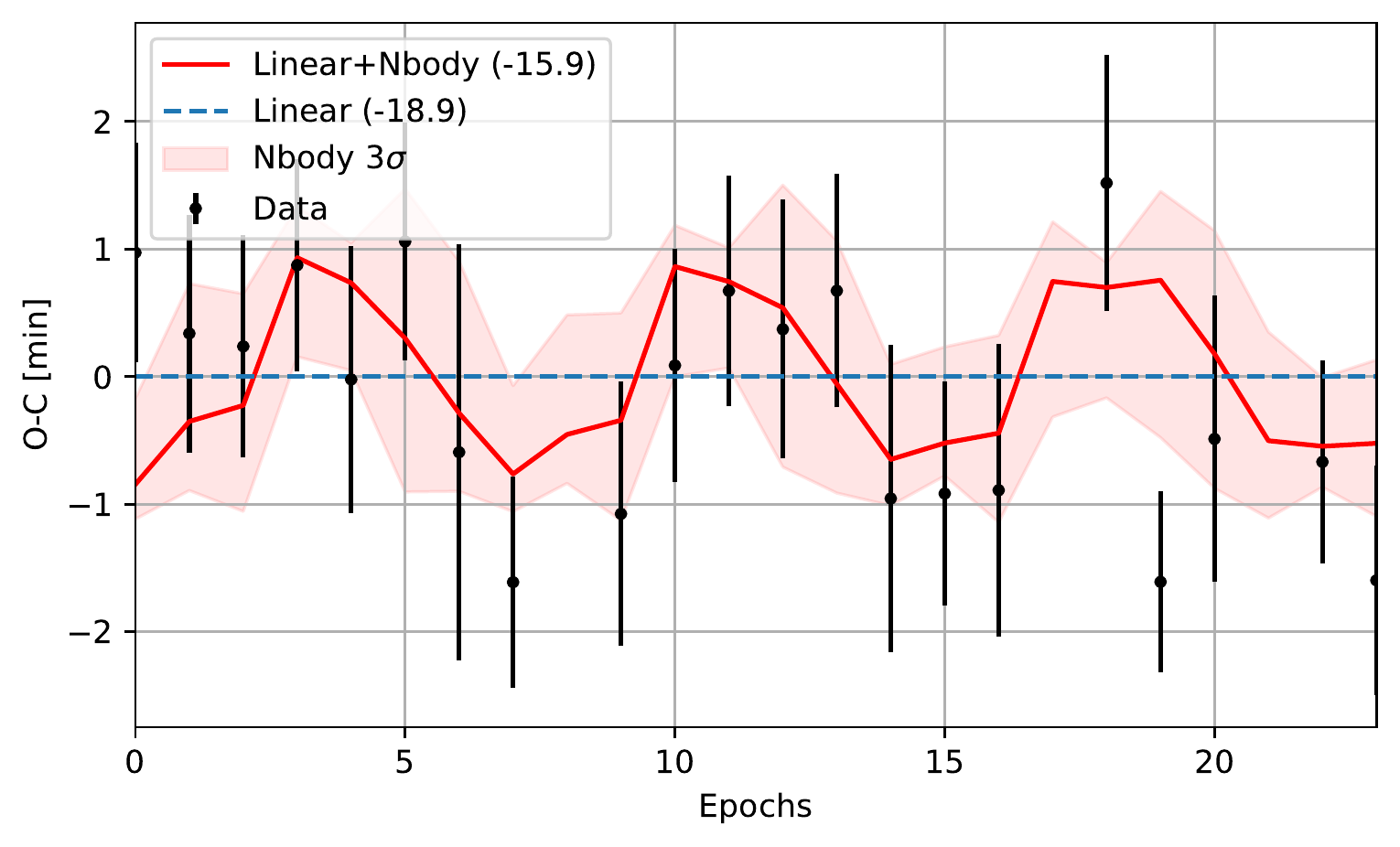}
\caption{ The residuals of a linear fit to the transit times are plotted and then compared against a non-linear model computed from an N-body simulation. The values in the legend indicate the Bayesian evidence output from MultiNest. The non-linear model has a better evidence (-15.9) than the linear model (-18.9). The red shaded region indicates the variability in the calculated perturbation as a result of the uncertainties in the derived system parameters in Table \ref{tab:ephem}. }
\label{wasp126_model}
\end{figure}

% The distributions shown here marginalize the argument of periastron. The argument of periastron is derived separately for each parameter space evaluation and chosen to best fit the data. Optimizing the nested sampling algorithm in this way reduces the computational requirements from $\sim$10 hours per retrieval to $\sim$1.

% \begin{figure}[H]
% \includegraphics[scale=0.30]{wasp_126bc_rv.png}
% \caption{ A comparison of the modes from the posteriors in Figure \ref{nested_post} showing the difference in RV models for the WASP-126 system. The RV measurements are from \cite{Maxted2016} and phase folded to the period of WASP-126 b, 3.28 day. The model in green corresponds to a Jupiter mass object at a period of 11 days while the red model represents a 69 Earth mass object at 7.5 day period. The scatter in the RV data suggests the smaller planet model is a better fit since a more massive planet would introduce more variability into the RV signal than what is already seen. }
% \label{wasp126_rv}
% \end{figure}

\subsection{WASP-18}

\begin{figure}[H]
\textbf{Time series analysis of WASP-18} 

\includegraphics[scale=0.4]{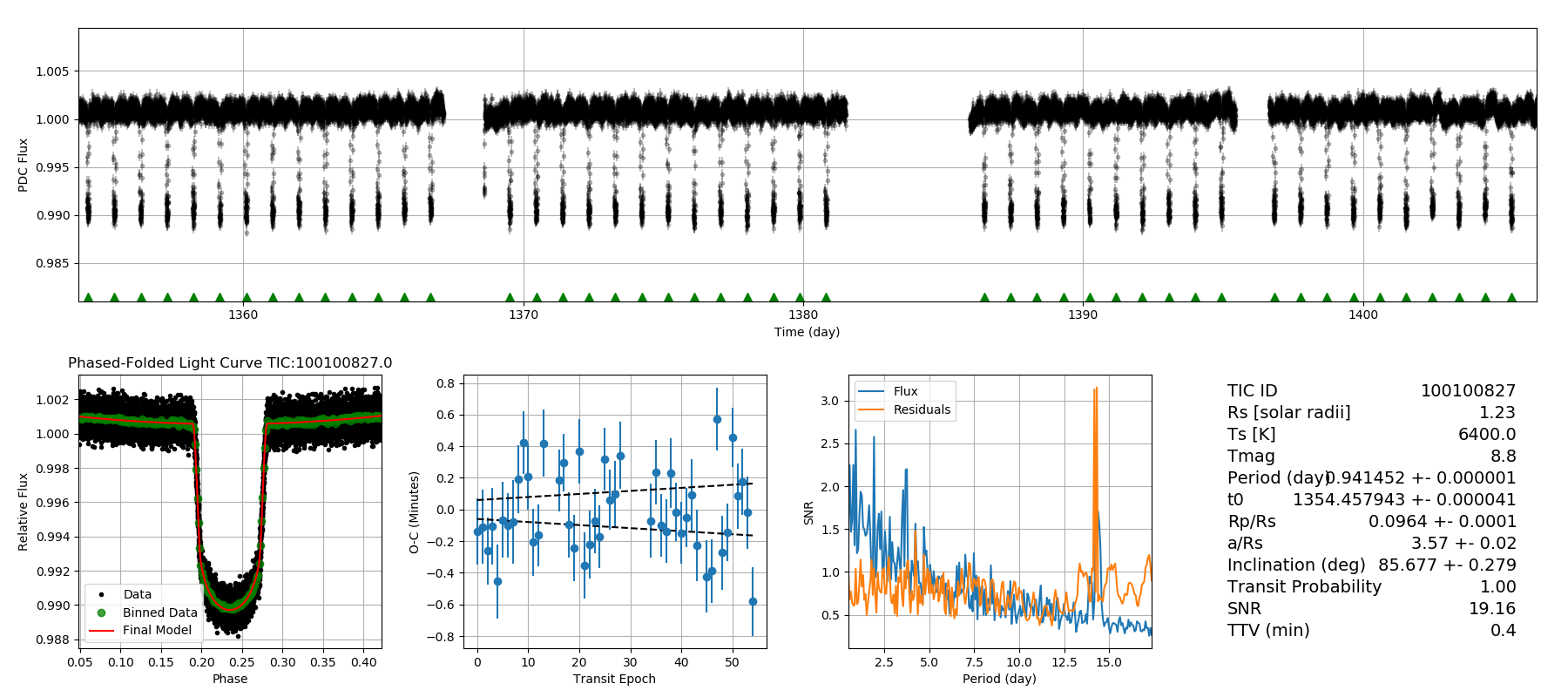}
\caption{ A time series analysis for the object WASP-18 b using the data from TESS. The top subplot shows a full time series for data from sectors 1--3 with the mid transit of each light curve plotted as a green triangle. The bottom left subplot shows a phase folded light curve that has been fit with a transit model to derive the planetary parameters shown in the table on the far right. The green data points are phase folded and binned to a cadence of 2 minutes. The middle subplot shows the residuals of a linear ephemeris (calculated) fit to the mid-transit times. The dotted line in the O-C plot represents one sigma uncertainties on the linear ephemeris. The middle right subplot shows a transit periodogram for the PDC Flux and for the residuals of the time series, after each light curve signal is removed. The spike at 13.8 days in the transit periodogram results from the gaps in the time series due to space craft maneuvers. }
\label{timeseries_wasp18}
\end{figure}

WASP-18 is massive (10.4 M$_{Jup}$) Jupiter-sized planet orbiting a 6400 K F type star at an orbital period of 0.94 day (\cite{Hellier2009}; \cite{Southworth2009}). WASP-18 b is an interesting candidate for studying tidal interactions and star-planet interactions because of its short orbital period and extreme mass (\cite{Csizmadia2019}; \cite{Barker2010}; \cite{Arras2012}; \cite{Miller2012}; \cite{Albrecht2012}; \cite{Wilkins2017}). WASP-18 b was observed with Spitzer and found to have a brightness temperature of $\sim$3100 K which yields a secondary eclipse depth of 0.38$\%$ (\cite{Nymeyer2011}; \cite{Maxted2013}). Additionally, the transit times from \cite{Maxted2013} show deviations on the order of $\sim$60 seconds, the same level as those seen here with TESS. Recently, a full-orbit phase curve in the optical was measured with TESS \citep{Shporer2018}. The optical phase curve yielded a low albedo, lack of atmospheric phase shift and inefficient heat distribution from day to night hemispheres. The phase curve variations affect the measurement of the transit parameters since it adds a non-linear component into the baseline (e.g. see the phased-folded data in Figure \ref{timeseries_wasp18}). As opposed to modeling the phase curve, which is beyond the scope of this project, the non-linearity is accounted for by simultaneously fitting a quadratic function along with the light curve model (see Equation \ref{lcmodel}). This approach yields a transit depth of 9293 $\pm$ 21 ppm which is inconsistent to 4.26 $\sigma$ with the 9439$^{27}_{26}$ ppm derived in \cite{Shporer2018}. The inconsistency is due to differences in limb darkening parameters considering the other parameters like inclination and a/R$_{s}$ are consistent. \textcolor{black}{A separate analysis with the same limb darkening parameters as \cite{Shporer2018} yielded results consistent within 0.5 $\sigma$.} 

\textcolor{black}{
The effects of contaminating sources can alter the transit depth of a planet by making it appear smaller than it actually is. \cite{Shporer2018} indicated two moderately bright stars in the TESS catalog that are within the vicinity of WASP-18, one of targets in this study. The stars TIC 100100823 and TIC 100100829 are 73`` and 83`` away from WASP-18 and have TESS magnitudes of T = 12.65 mag and T = 12.50 mag, respectively. Compared to WASP-18 (T = 8.83 mag) the nearby sources are 34 and 29 times fainter, respectively. \cite{Shporer2018} found the stars would contribute minimally (0.01$\%$) to the flux of WASP-18 so the same aperture size (that of SPOC) is adopted here as the former study on WASP-18 b's phase curve. 
}

The residuals of a linear ephemeris display transit timing variations on the order of 30 seconds with measurements being correlated in time. \cite{Shporer2018} also searched for TTVs in the WASP-18 system and concluded there was more evidence for no TTV. The findings here contradict their result because different models are used to assess the signal in the data. \cite{Shporer2018} use a parameterized sin wave to fit the residuals of their linear ephemeris. Here, the transit times themselves are fit with a more complicated model rather than to the residuals in a disjointed approach. The Bayesian evidence is then compared between a linear and non-linear ephemeris to assess the significance of another planet being present in the system. For consistency reasons with respect to the data acquisition and time keeping, past measurements of WASP-18 b are ignored and only the TESS measurements are used to assess the TTV signal. The N-body analysis uses an eccentricity value of 0 for WASP-18 b which is consistent with the value measured from Spitzer of 0.0091 $\pm$ 0.0012 \citep{Nymeyer2011}. The best fit companion from the retrieval has a mass of 55.2 $\pm$ 12.3 $M_{Earth}$ and a period of 2.155 $\pm$ 0.006 day. The global log evidence from MultiNest is reported in the legend of Figure \ref{ttv_explore_wasp18} and is greater by 10.9 for the non-linear ephemeris. Despite the close orbital distances and extreme mass of WASP-18b, the hill spheres of WASP-18 and the perturber do not interact. The hill sphere for WASP-18 b is 0.003 AU and it has an orbital distance of 0.02 AU, where as the companion has a hill radius of 0.0015 AU and an orbital distance of 0.035 AU. 

There is no evidence for additional transiting companions in the system based on a periodogram analysis. The periodogram shows a spike around 13.8 days consistent with the orbits of the spacecraft and some minor signals around 5 days, which are suspect to come from the star. There is a rotation period for the star measured around 5.6 days (\citealt{Hellier2009}; \citealt{Maxted2013}). The inclination limit for a transiting planet at 2.12 days is 80.6 degrees and is an upper limit for the true value. The companion would cause a RV signal of $\sim$18--29 m/s which is small compared to the $\sim$ 1800 m/s value for WASP-18 b. The companion's RV signal is interestingly close to the predicted amplitude of RV jitter due to tidal distortions in the host star from WAPS-18 b \citep{Arras2012}. \textcolor{black}{ It seems possible to detect a companion in this system considering the 9 RV measurements from \cite{Hellier2009} have an error of $\sim$ 10 m/s. The most recent analysis from \cite{Bonomo2017} reports a semi-amplitude of WASP-18 b to be 1817.2 $\pm$ 2.3 m/s with an upper limit on the RV jitter of the star to be 3.5 m/s. Additionally, the measurements in \cite{Triaud2010} exhibit scatter on the order of $\sim$10-25 m/s. However, it's difficult to say if the past measurements resolve the RV peak from the companion because of aliasing and whether they're consistent with the results here. In which case, it would be worth performing a joint simultaneous fit between the TTV signal and RV measurements in order to further constrain the companion's mass and orbit. The orbital constraints placed on the outer candidate here can be used to optimize follow-up RV measurements and prevent errors from aliasing the signal. }

\begin{figure}
\centering
\textbf{Posteriors for WASP-18 Non-linear Ephemeris}

\includegraphics[scale=0.5]{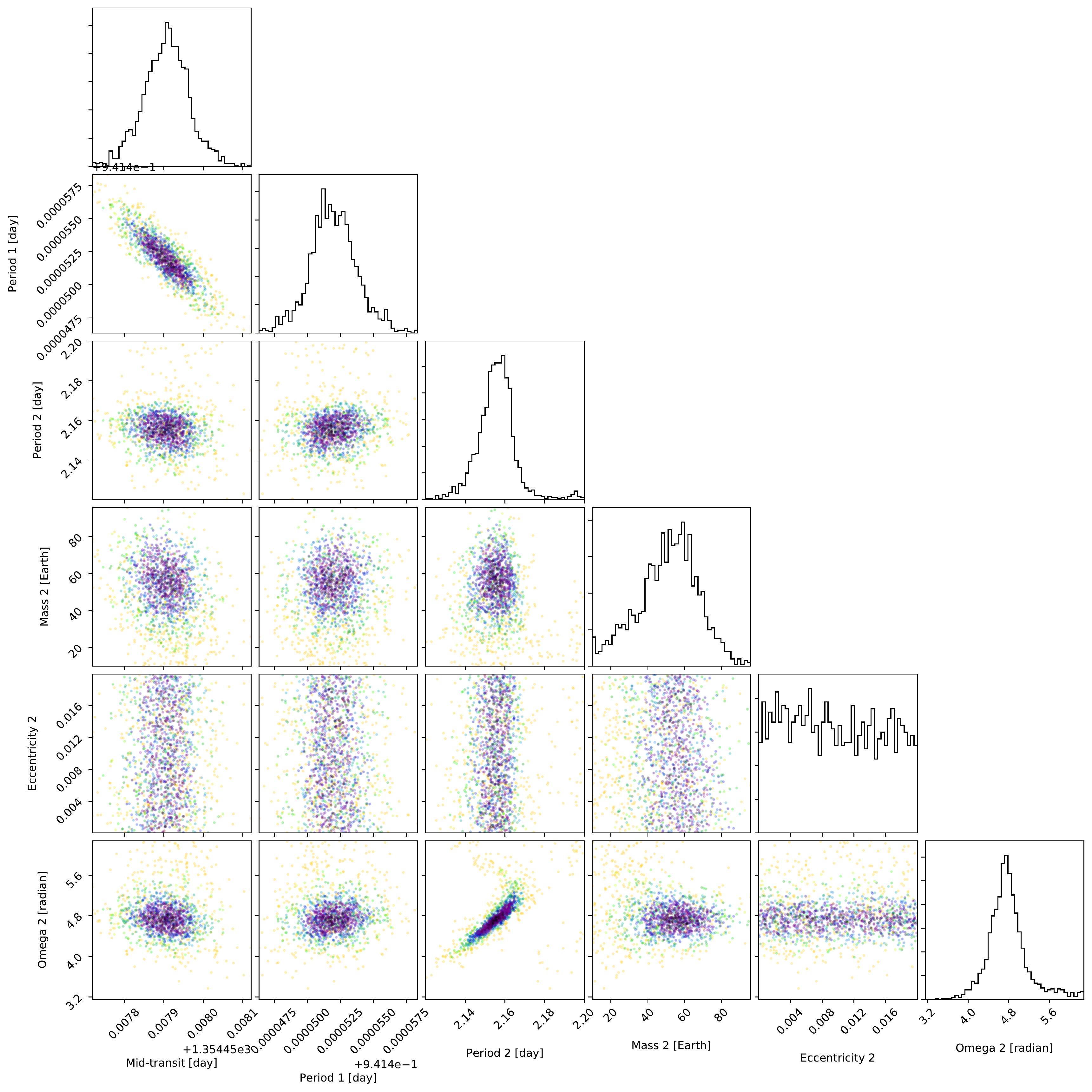}\newline
\caption{ Posteriors for the non-linear ephemeris model are plotted for the WASP-18 system. The color of the data points are coordinated to the fit percentile, with colors darker than yellow indicating fits better than the 50th percentile. The final parameters and uncertainties are in Table \ref{tab:ephem}.
}
\label{ttv_explore_wasp18}
\end{figure}

\begin{figure}
\centering
\textbf{WASP-18 Ephemeris Comparison}

\includegraphics[scale=0.75]{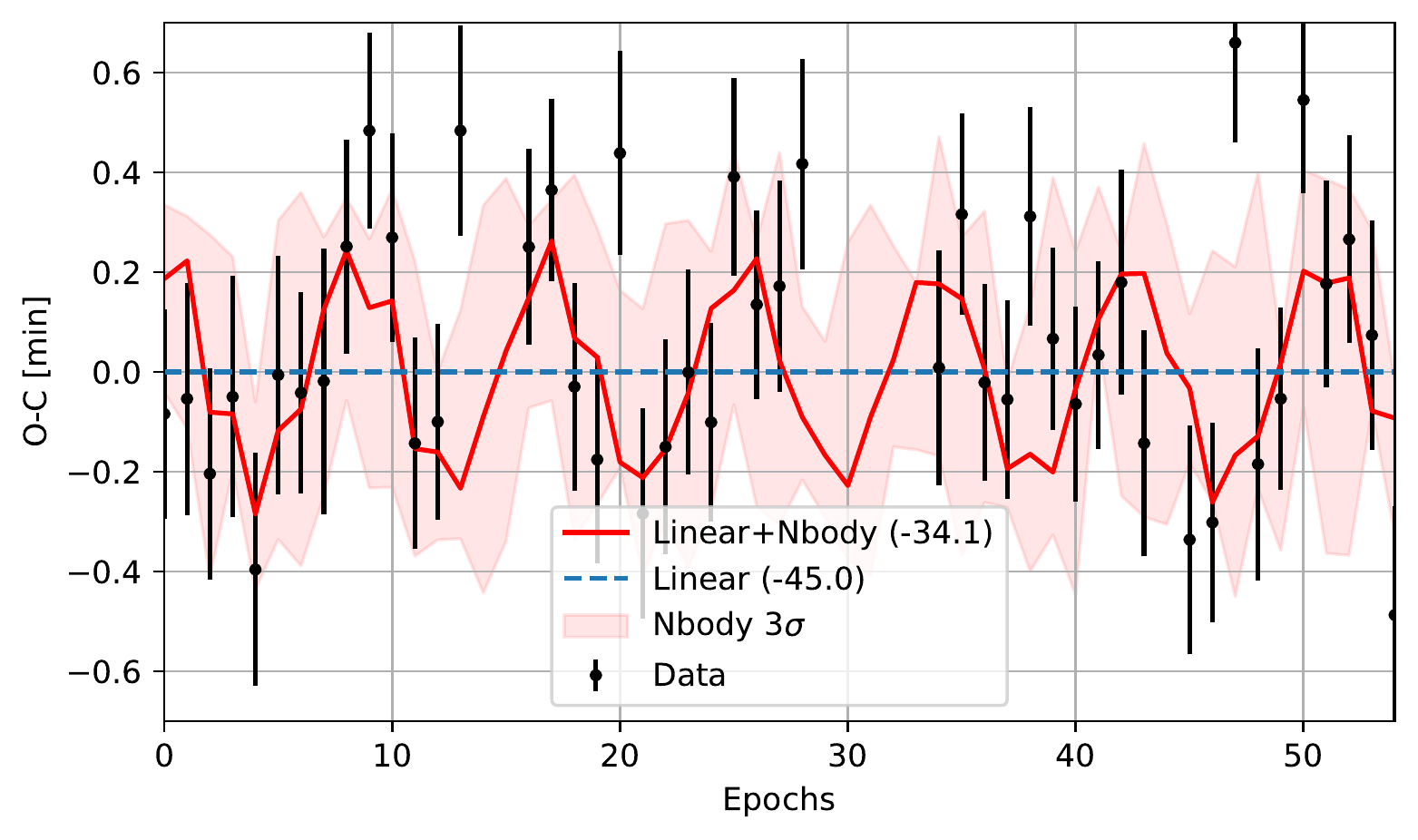}
\caption{ The residuals of a linear ephemeris are plotted and then compared against a non-linear model computed from an N-body simulation. The values in the legend indicate the Bayesian evidence output from MultiNest. The non-linear model has a greater evidence than the linear model suggesting the perturbation model is a better fit to the data. The red shaded region indicates the variability in the TTV signal from uncertainties in the derived parameters in Table \ref{tab:ephem}. }
\label{ttv_explore_wasp18}
\end{figure}

\subsection{TOI 193}

\begin{figure}
\includegraphics[scale=0.4]{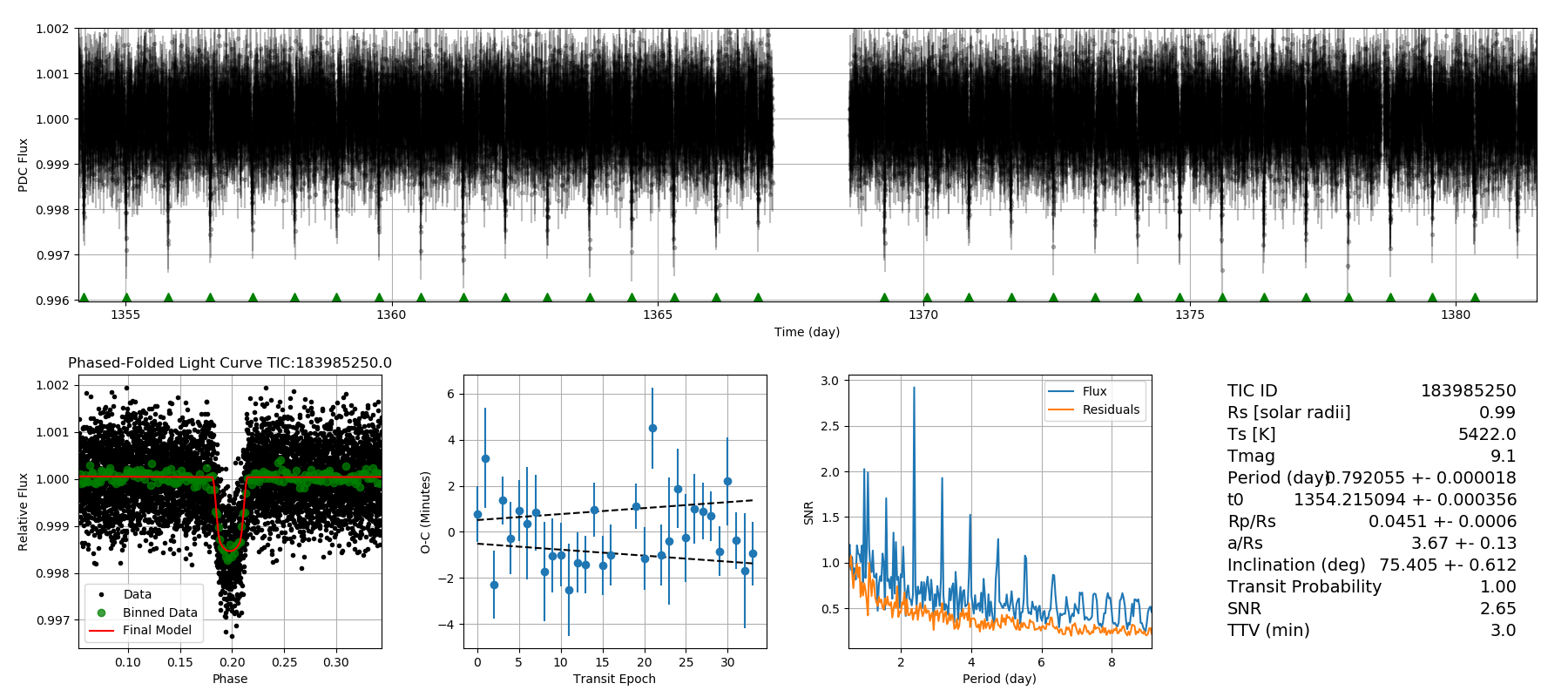}
\caption{ A time series analysis for the object TOI 193.01 using the data from TESS. The top subplot shows a full time series for data from sectors 1--2 with the mid transit of each light curve plotted as a green triangle. The bottom left subplot shows a phase folded light curve that has been fit with a transit model to derive the planetary parameters shown in the table on the far right. The green data points are phase folded and binned to a cadence of 2 minutes. The structure of the residuals is analyzed in Figure \ref{ttv_model_toi193}. The dotted line in the O-C plot represents one sigma uncertainties on the linear ephemeris. The middle right subplot shows a transit periodogram for the PDC Flux and for the residuals of the time series, after each light curve has been removed. }
\label{timeseries:toi193}
\end{figure}

The last object of interest is a candidate for a multi-planet system, TOI 193 or TIC 183985250. TOI 193 is a G type star with a radius of 0.99 R$_{Sun}$. Since TOI 193 is an object of interest to the TESS community and active follow-up is being conducted to constrain the system properties a few assumptions are made. For the N-body simulation the star is assumed to have a mass of 1 M$_{Sun}$ and the transiting planet is assumed to have an eccentricity of 0. \textcolor{black}{The phase folded light curve yields a radius of 4.9 R$_{Earth}$ which corresponds to a mass of $\sim$22 M$_{Earth}$ based on the Gaussian process in Figure \ref{mass_radius}.} The detection of the transit was a little more tenuous then the others in this study because it has an SNR of 2.65. The time series analysis is shown in Figure \ref{timeseries:toi193}. \textcolor{black}{A search for contaminating light sources around TOI 193 (TIC 183985250) yielded one star within 60`` (about the photometric aperture size). The neighboring star (TIC 183985251) is 8.2`` away with a magnitude of T = 13.957. The star is 88 times dimmer but given the distance to the target it corresponds to $\sim$1.1$\%$ of the flux in the aperture. The average displacement in x and y of the target TOI 193 is 0.025 pixels with the max being 0.11 pixels. The contaminating star likely has a similar point error and is going to be in the same pixel as the target most of the time. Therefore shifts on the CCD are unlikely to induce the transit signal seen because the contaminating star would have to drift out of the aperture and back in within a transit duration. }

\begin{figure}
\centering
\textbf{Posteriors for TOI 193 Non-linear Ephemeris}

\includegraphics[scale=0.5]{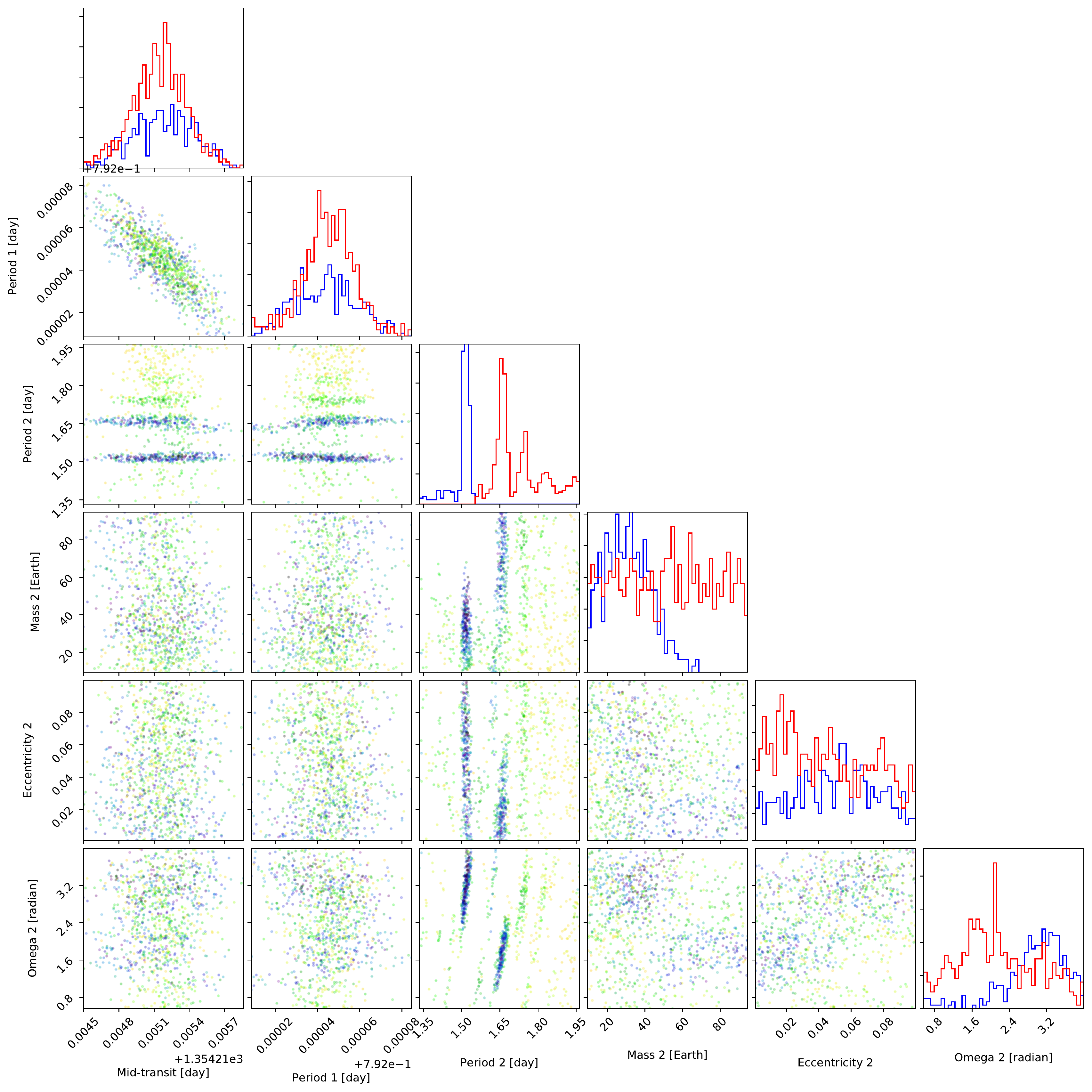}\newline
\caption{ 
 Posteriors for the non-linear ephemeris model are plotted for the TOI-193 system. The color of the data points are coordinated to the fit percentile, with colors darker than yellow indicating fits better than the 50th percentile. The final parameters and uncertainties are in Table \ref{tab:ephem}. The colored histograms represent two possible solutions that are degenerate and separated based on the period of the perturbing body. Each mode is plotted in their respective color in Figure \ref{ttv_model_toi193}. 
 }
\label{ttv_posterior_toi193}
\end{figure}

\begin{figure}
\includegraphics[scale=0.75]{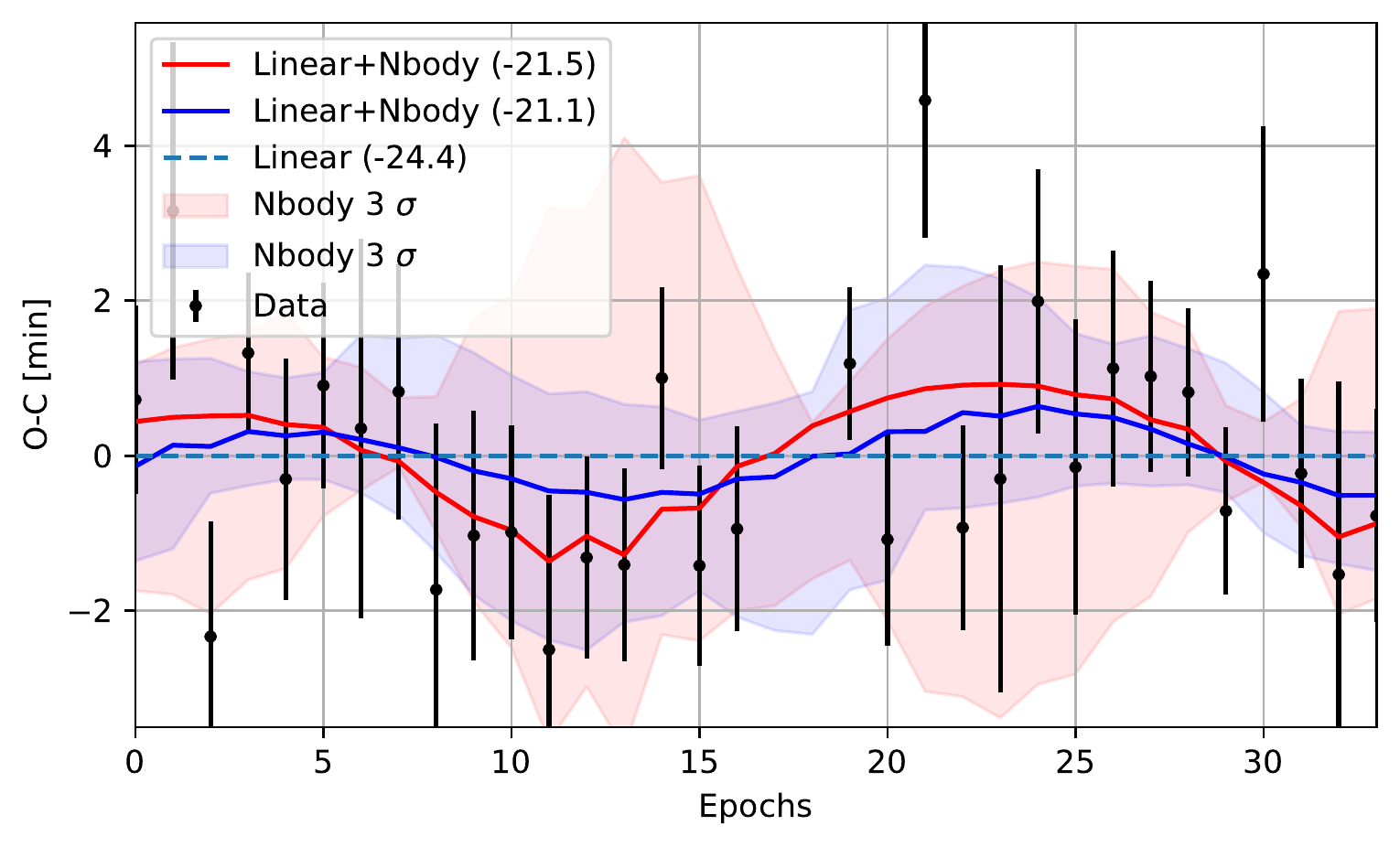}
\hspace{-0.4in}
\caption{ The residuals of the linear ephemeris are plotted and then compared against the non-linear models computed from N-body simulations. The values in the legend indicate the Bayesian evidence output from MultiNest. The non-linear models are degenerate as indicated by the two modes in the posteriors. The N-body range is computed from the min and max of multiple N-body models with different combinations of parameters between +/- 3 sigma. The color of each model represents a degenerate solution set from the posteriors in Figure \ref{ttv_posterior_toi193}. 
%The degeneracy lies with adding eccentricity as a free parameter into the retrieval. The two modes yield similar chi-squared values and have evidence values greater than a linear ephemeris. The two possible modes are a companion with 45.1 M$_{Earth}$, 1.52 day period, 0 eccentricity and 58.3 M$_{Earth}$ at 1.65 day period and eccentricity of 0.03. The red shaded region indicates the variability in the TTV signal as a result of the uncertainties in the derived parameters. 
}
\label{ttv_model_toi193}
\end{figure}

There is more evidence for the non-linear ephemeris compared to a linear ephemeris however the non-linear ephemeris is degenerate. The final parameters for each ephemeris are shown in Table \ref{tab:ephem}. Two degenerate TTV modes are plotted in the posteriors (Figure \ref{ttv_posterior_toi193}) because each solution has a Bayesian evidence greater than the linear ephemeris. Each perturbation solution is close to the orbital resonance of 2:1. The two solutions look similar (see Figure \ref{ttv_model_toi193}) but yield noticeable different RV signals. Future RV observations will be able distinguish the correct mode because the companion can cause as much variation in the RV semi-amplitude as 54 m/s but as little as $\sim$15, depending on the mass. There is no evidence for a transiting companion at an SNR greater than 1 in the residuals of the time series. The periodic structure in the transit periodogram is an aliasing effect of the short period transit being phase folded into the data. 

% The distribution only shows mass and period with the omega value being marginalized during the retrieval by deriving the optimal value for each location in the constrained parameter space. This increases the clarity of degeneracies in the posteriors by ignoring values that do not contribute significantly to the signal. 

\begin{table}[H]
  \caption{Ephemeris Parameters}
  \label{tab:ephem}
  \begin{center}
\begin{tabular}{lllll} \hline \hline
& WASP-18 b & WASP-126 b & TOI 193 & TOI 193  \\
\hline
TIC ID          & 100100827     & 25155310 & 183985250  & 183985250    \\ 
M$_{*}$ [M$_{.}$] &1.22 &1.12 &1 &1 \\
M$_{1}$ [M$_{Earth}$] &3314.82 &90.3 &22.0 &22.0 \\

&&& \\ 
Linear Evidence & -45.0 & -18.9 & -24.4 & -24.4   \\
L. Period [day] & 0.941452 $\pm$ 1.34e-06  & 3.288782 $\pm$ 1.94e-05 & 0.792043 $\pm$ 1.82e-05 & 0.792043 $\pm$ 1.82e-05  \\
L. T$_{mid}$    & 1354.457943 $\pm$ 4.15e-05 & 1327.520692 $\pm$ 2.57e-04 & 1354.215278 $\pm$ 3.54e-04 & 1354.215278 $\pm$ 3.54e-04  \\
& & & \\ 
Non-linear Evidence     & -34.1                 & -15.9                     & -21.1                     & -21.5 \\
NL. P$_{1}$ [day] & 0.941240 $\pm$ 2.34e-06     & 3.28883 $\pm$ 3.71e-05    & 0.792074 $\pm$ 2.39e-05   & 0.792169 $\pm$ 3.44e-05    \\
NL. T$_{mid}$    & 1354.458393 $\pm$ 5.45e-05   & 1327.5203 $\pm$ 3.85e04   & 1354.216555$\pm$ 6.60e-04 & 1354.215943$\pm$ 8.09e-04 \\
NL. Mass 2 [M$_{Earth}$] & 55.2$\pm$12.3        & 64.2$\pm$24.6             & 39.4$\pm$9.5              & $<$ 100         \\ 
NL. Period 2 [day]       & 2.1558$\pm$0.0064    & 7.63$\pm$0.17             & 1.516$\pm$0.021           & 1.65$\pm$0.04         \\ 
NL. Eccentricity 2       & $<$0.015      & $<$0.05                   & $<$ 0.02                  & $<$ 0.1       \\
NL. Omega 2 [rad]        & 4.74$\pm$0.24        & 3.197$\pm$0.59            & 3.19$\pm$0.51             & 1.92$\pm$0.78         \\ 
\end{tabular}
\end{center}
\end{table}

%\begin{table}[H]
%  \caption{Ephemeris Parameters}
%  \label{tab:ephem}
%  \begin{center}
%\begin{tabular}{lllll} \hline \hline
%& WASP-18 b & WASP-126 b & TOI 193 & TOI 193  \\
%\hline
%TIC ID          & 100100827     & 25155310 & 183985250  & 183985250    \\ 
%M$_{*}$ [M$_{.}$] &1.22 &1.12 &1 &1 \\
%M$_{1}$ [M$_{Earth}$] &3314.82 &90.3 &22.0 &22.0 \\

%&&& \\ 
%Linear Evidence & -45.0 & -18.9 & -24.4 & -24.4   \\
%L. Period [day] & 0.941452 $\pm$ 1.34e-06  & 3.288782 $\pm$ 1.94e-05 & 0.792043 $\pm$ 1.82e-05 & 0.792043 $\pm$ 1.82e-05  \\
%L. T$_{mid}$    & 1354.457943 $\pm$ 4.15e-05 & 1327.520692 $\pm$ 2.57e-04 & 1354.215278 $\pm$ 3.54e-04 & 1354.215278 $\pm$ 3.54e-04  \\
%& & & \\ 
%Non-linear Evidence     & -34.1                 & -15.9                     & -21.1                   & -21.5 \\
%NL. P$_{1}$ [day] & 0.941240 $\pm$ 2.34e-06     & 3.28883 $\pm$ 3.16e-05    & 0.792074 $\pm$ 2.39e-05 & 0.792169 $\pm$ 3.44e-05    \\
%NL. T$_{mid}$    & 1354.458393 $\pm$ 5.45e-05   & 1327.5202 $\pm$ 3.09e04   & 1354.216555$\pm$ 6.60e-04& 1354.215943$\pm$ 8.09e-04 \\
%NL. Mass 2 [M$_{Earth}$] & 55.2$\pm$12.3        & 64.0$\pm$26.9             & 39.4$\pm$9.5      & 73.4$\pm$28.1         \\ 
%NL. Period 2 [day]       & 2.1558$\pm$0.0064    & 7.65$\pm$0.27             & 1.516$\pm$0.021   & 1.65$\pm$0.04         \\ 
%NL. Eccentricity 2       & 0.009$\pm$0.006      & 0.025$\pm$0.018           & 0.059$\pm$0.026   & 0.029$\pm$0.026       \\
%NL. Omega 2 [rad]        & 4.74$\pm$0.24        & 3.328$\pm$0.77            & 3.19$\pm$0.51     & 1.92$\pm$0.78         \\ 
%\end{tabular}
%\end{center}
%\end{table}

\section{Conclusion} 

The Transiting Exoplanet Survey Satellite (TESS) objects of interest (TOI) are analyzed in search of transit time variations. Data from sectors 1-3 are used to analyze 74 planet candidates from the TOI catalog. The average photometric precision of candidate hosting stars are measured. Stars with TESS magnitudes between $\sim$9--11 exhibit noise on the order of $\sim$1000 ppm. The extensive TESS candidate list of over 3 million targets requires an efficient analysis in order to produce results in a timely manner which can be used to refine follow-up projects (e.g. Exoplanet Transit Survey; \cite{Zellem2019}). Artificial intelligence is used to vet a portion of non-transit signals in order to expedite the characterization of numerous transits and ephemerides. A convolutional neural network is trained on time series measurements from TESS and has a transit detection accuracy of 99$\%$ with transits over an SNR of 2. Preliminary excavations using the AI vetting algorithm on the TESS \textcolor{black}{candidate target list (CTL)} yielded a large amount of eclipsing binary signals. Means to distinguish the eclipsing binaries from transit signals will be implemented in future versions of the AI. After characterizing individual transit measurements, residuals of a linear ephemeris are used to search for transit timing variations. To assess the significance of a perturbing planet the Bayesian evidence is compared between a linear and a non-linear ephemeris, which is based on an N-body simulation. There is evidence for 3 new multi-planet systems with non-transiting companions using the 2 minute cadence observations from TESS. Two previously known exoplanets, WASP-18 b and WASP-126 b are studied and there is evidence for WASP-18 c at a period of 2.155 $\pm$ 0.006 day with a mass of 55.2 $\pm$ 12.3 M$_{Earth}$ and WASP-126 c a 64.2 $\pm$24.6 M$_{Earth}$ planet at an orbital period of 7.63$\pm$0.17 day. The third system with a significant TTV is TOI 193.01 with a candidate planet, TOI 193.02, exhibiting a degenerate solution near the 2:1 resonance with a mass of either 39.4$\pm$9.5 or 73.4$\pm$28.1. Radial velocity observations in the future will be able to distinguish between the degenerate solutions because the RV semi-amplitude will vary between at least $\sim$15--54 m/s depending on the perturbing planet's mass. Results for these targets will be uploaded to the exoplanet TESS follow up program to encourage further analysis and collaboration with the community. TESS provides us with an exciting opportunity to perform high cadence observations of many new and existing exoplanet systems during the all sky survey. 

% Some stars were sensitive to subpixel shifts in the detector (e.g. Pi Mensae b) and determining the optimal aperture was necessary to reduce systematic effects. In the future, an analysis similar to the intra-pixel sensitivities of Spitzer might want to be looked into (e.g. using pixel level decorrelations CITE or independent component analysis CITE). In the future having follow-up for these system is crucial to maintain an accurate calendar. Follow up programs like cite zellem2019. 

% Finally, we estimate TTV perturbations for all known transiting multi-planet systems and find 42 targets with TTVs observable by TESS that could benefit from future high cadence follow-up in order to constrain the ephemerides and planet mass. WASP-126 b shows a subtle transit timing variation that would be indicative of a 66.2 Mearth planet at an orbital period of 7.49, close to the 2:1 resonance. WASP-126 b is fluffy planet with a radius of 0.95 R$_Jup$ around a moderately bright star (10.8 V mag), making it a good target for transmission spectroscopy follow-up.

% conducted a search through TOI catalog for TTV measurements
% Find statistical evidence for 3 new multiplanet systems 
% Measure photometric precision of TESS to be XX 
% Also find a few TOIs that did not pass our transit vetting algorithm 

\section{Data Availability}
Photometric data from the Transiting Exoplanet Survey Satellite is available online: \url{https://archive.stsci.edu/tess/bulk_downloads.html}. 

\section{Code Availability}
The machine learning algorithm used to vet transit candidates is available here: \url{https://github.com/pearsonkyle/Exoplanet-Artificial-Intelligence}. The algorithm used to model exoplanet transits with nested sampling is available online: \url{https://github.com/pearsonkyle/Exoplanet-Light-Curve-Analysis}. The N-body retrieval algorithm is available online: \url{https://github.com/pearsonkyle/Nbody-AI}.

\section{Acknowledgements}
I would like to thank my advisor, Caitlin Griffith, for supplying me with my first research assistantship (RA) so I could focus on writing my thesis proposal even though most of my time was spent on this study. I thank Eli Shneider and the whole Bently's crew for fueling parts of this research and for providing me with a location to work when campus was too distracting. Daynal Tansu and Avi Shporer for their discussion about WASP-18 b. Adam Sutherland for putting up with my questions about orbital dynamics while on the ski lift. Last but not least, the anonymous referee for providing helpful comments which strengthened the paper. 
\bibliographystyle{aasjournal}
\bibliography{ref}
\end{document}